\documentclass{aa}  

\usepackage{graphicx}
\usepackage{txfonts}
\usepackage{subcaption}
\usepackage{lscape}
\usepackage{placeins}
\usepackage{natbib}
\usepackage{xcolor}
\newcommand{\cedit}[1]{#1}
\newcommand{\redit}[1]{#1}
\let\linenumbers\relax
\usepackage[colorlinks=true,citecolor=blue,linkcolor=blue,urlcolor=blue]{hyperref}
                                
\begin{document}

   \title{UNIONS: Red and blue alignments of DESI DR1 galaxies}


%

   \author{R. Paviot\inst{1,2}
           \and C. Murray\inst{1}
           \and M. Kilbinger\inst{1}
           \and C. Daley\inst{1}
           \and F. Hervas-Peters\inst{1,3}
           \and A. Corinaldi\inst{1}
           \and T. de Boer\inst{4}
           \and S. Guerrini\inst{1}
           \and M.~J.~Hudson\inst{5,6,7}
           \and G.~S.~H.~Paek\inst{4}
           \and L. Van Waerbeke\inst{8}
           \and J. Yun\inst{9,10}
          }

   \authorrunning{R. Paviot et al.}

   \institute{Universit\'e Paris-Saclay, Universit\'e Paris Cit\'e, CEA, CNRS, AIM, 91191, Gif-sur-Yvette, France
         \and
             Centre National d'\'Etudes Spatiales -- Centre spatial de Toulouse, 18 avenue Edouard Belin, 31401 Toulouse Cedex 9, France
         \and
             Steward Observatory, University of Arizona, 933 North Cherry Avenue, Tucson, AZ 85721, USA
         \and
             Institute for Astronomy, University of Hawai'i, 2680 Woodlawn Drive, Honolulu, HI 96822, USA
         \and
             Department of Physics and Astronomy, University of Waterloo, 200 University Avenue West, Waterloo, Ontario N2L 3G1, Canada
         \and
             Waterloo Centre for Astrophysics, University of Waterloo, Waterloo, Ontario N2L 3G1, Canada
         \and
             Perimeter Institute for Theoretical Physics, 31 Caroline St. North, Waterloo, ON N2L 2Y5, Canada
         \and
             Department of Physics and Astronomy, University of British Columbia, 6224 Agricultural Road, Vancouver, BC V6T 1Z1, Canada
         \and
             Department of Astronomy, Graduate School of Science, The University of Tokyo, 7-3-1 Hongo, Bunkyo-ku, Tokyo 113-0033, Japan
         \and
             Institute of Astronomy, Graduate School of Science, The University of Tokyo, 2-21-1 Osawa, Mitaka, Tokyo 181-0015, Japan
             }

   \date{Received XXX; accepted XXX}

 
  \abstract
   {\redit{Intrinsic alignments (IA) of galaxies are a dominant source of
   systematics in weak-lensing cosmological analyses, and direct measurements on
   spectroscopic samples are the primary source of priors on their amplitude.}}
   {\redit{We measure the IA of galaxies from the first data release (DR1)
   of the Dark Energy Spectroscopic Instrument (DESI) as a function of colour,
   stellar mass, and halo mass, by cross-correlating the DESI spectroscopic
   catalogues with the weak-lensing shape catalogue of the Ultraviolet
   Near-Infrared Optical Northern Survey (UNIONS) over up to
   $2400\,{\rm deg}^2$.}}
   {\redit{We fit the three-dimensional galaxy--galaxy and galaxy--shape
   multipoles of the DESI Bright Galaxy Survey (BGS), luminous red galaxy
   (LRG), and emission-line galaxy (ELG) samples, split by colour and stellar
   mass, with the non-linear linear alignment (NLA) and the tidal alignment and
   tidal torquing (TATT) models. Halo masses are measured for the BGS red
   sample from galaxy--galaxy lensing and extended to the LRGs through the
   resulting stellar-to-halo mass relation.}}
   {\redit{Red galaxies are strongly and radially aligned, but their
   alignment amplitude is not a single power law of halo mass between
   $6\times10^{11}$ and $3\times10^{13}\,h^{-1}M_\odot$: it is nearly flat at
   low mass and rises steeply above a break at
   $1.3\times10^{13}\,h^{-1}M_\odot$. Massive blue BGS galaxies show instead
   a tangential alignment, detected at $3.3\sigma$, to our knowledge the most
   significant detection of a non-zero alignment of blue galaxies to date,
   while the low-mass blue BGS galaxies and the ELGs are consistent with zero.
   At our current precision, the data do not favour TATT over NLA.}}
   {\redit{These measurements extend direct IA constraints towards the fainter
   and bluer galaxies that dominate cosmic-shear samples. The flat low-mass
   end of the red population has to be accounted for when extrapolating IA
   priors to fainter galaxies. DESI DR2 will allow these fits to be restricted
   to larger scales, where the NLA and TATT models are more reliable.}}


   \maketitle

\section{Introduction}
\label{sec:introduction}

Cosmic shear, the coherent distortion of \redit{the images of background galaxies} by the gravitational lensing effect of large-scale structure, has emerged as one of the most powerful probes of cosmology. Recent stage-III surveys --- the Dark Energy Survey \citep[DES;][]{Amon21,DESY6}, the Kilo-Degree Survey \citep[KiDS;][]{Heynmans2021,NLA-M}, and the Hyper Suprime-Cam survey \citep[HSC;][]{Dalal2023,Li2023}\redit{, and the Ultraviolet Near-Infrared Optical Northern Survey \citep[UNIONS;][]{Gwyn2025,Goh2026,Guerrini2026}} --- have demonstrated that weak lensing measurements can constrain the amplitude of matter fluctuations \redit{to a precision of a few per cent}. The upcoming stage-IV surveys, \textit{Euclid} \citep{Mellier2024}, the Vera C.\ Rubin Observatory Legacy Survey of Space and Time \citep[LSST;][]{Ivezic2019}, and the \textit{Nancy Grace Roman Space Telescope} \citep{Eifler2021}, will observe billions of galaxies, making previously sub-dominant systematic effects significant.

Among the most important systematics for cosmic shear is the intrinsic alignment (IA) of galaxies. As galaxies form and evolve within the cosmic web, tidal forces from the surrounding large-scale structure induce coherent orientations in their shapes \citep{Catelan2001,HirataSeljak2004,Crittenden01}. These intrinsic shape correlations mimic or contaminate the cosmic shear signal in two ways: through correlations between the intrinsic shapes of physically nearby galaxies (II correlations) and through correlations between the intrinsic shape of a foreground galaxy and the gravitational shear of a background galaxy (GI correlations). If not properly accounted for, intrinsic alignments can bias cosmological parameter inference from weak lensing surveys \citep[for reviews, see][]{joachimi_galaxy_2015,TroxelIshak2015,2024OJAp....7E..14L,Chisari2025review}.

Considerable observational effort has been devoted to characterising intrinsic alignments through direct measurements. A strong\redit{, radial} IA signal has been detected for red, pressure-supported galaxies, whose shapes tend to stretch toward matter overdensities \citep{Joachimi2011,Singh2015,Johnston2019,Fortuna21,Samuroff2023,HervasPeters2024}. The amplitude of this signal depends on luminosity, following a broken power law with a steep rise for bright galaxies \citep{Fortuna21,HervasPeters2024}, while \cite{Fortuna2025} \redit{suggested} that this broken power law behaviour was a consequence of the break in the stellar-to-halo mass relation; with halo mass being a more robust proxy of the IA signal. Blue, rotationally-supported galaxies, on the other hand, are expected to acquire \redit{spins correlated with} the local tidal field via tidal torquing \citep{Schafer2009,Codis18}. Observationally, blue galaxies show alignment signals consistent with zero at low and intermediate redshift \citep{mandelbaum_wigglez_2011,Johnston2019,Samuroff2019,Samuroff2023,HervasPeters2024,David2025}, though hydrodynamical simulations suggest a non-negligible signal may exist, especially at higher redshift \redit{\citep{Chisari15,Tenneti2015,Codis18,Samuroff21,Zjupa2022,Delgado2023}}.

\redit{Several perturbative models have been developed to describe intrinsic alignments in cosmological analyses.} The \redit{non-linear linear alignment} model \citep[NLA;][]{Bridle07,HirataSeljak2004} extends the linear alignment model of \citet{Catelan2001} by replacing the linear matter power spectrum with its non-linear counterpart\redit{, an empirical prescription rather than a result of perturbation theory}. The tidal alignment and tidal torque model \citep[TATT;][]{Blazek15,Blazek19} further incorporates higher-order terms, including \redit{the quadratic tidal-torquing contribution that theory predicts for disc galaxies}. \redit{The halo-mass picture of \citet{Fortuna2025} has motivated mass-dependent variants of both models: \citet{NLA-M} adopted NLA-M in the KiDS-Legacy cosmic shear analysis, parametrising the IA amplitude as a function of halo mass, and \citet{Herle2026} introduced TATT-M, calibrated on the FLAMINGO hydrodynamical simulations, which is strongly preferred over the standard NLA model.} More recently, effective field theory (EFT) approaches have been proposed to extend the model validity to smaller scales \citep{Vlah2019,Vlah2020,Bakx2023,Maion2023}. \cedit{NLA and TATT are already marginalised over in cosmological analyses, while the EFT models have so far \redit{mostly been validated on simulations, with a first application to cosmic-shear data by \citet{Chen2026}}.}

To date, however, most direct IA measurements have focused on bright, red galaxies selected from spectroscopic surveys. This focus on luminous red objects introduces a demographic bias in our understanding of intrinsic alignments, leaving the IA signal of fainter and bluer populations --- which dominate the source samples of cosmic shear surveys --- poorly constrained. \redit{The colour dependence of IA matters for cosmic shear itself: early-type galaxies dominate the IA contamination \citep{Heymans2013}, colour-split source samples have served as internal-consistency tests of cosmic-shear analyses \citep{Stolzner2025}, and restricting the sources to blue galaxies, as proposed by \citet{Krause2016} and done for DES Y3 \citep{McCullough2026}, reduces the sensitivity of the cosmological constraints to the IA model. These analyses still marginalise over the alignment of the blue galaxies, so direct measurements of it provide the priors needed to apply such selections in \textit{Euclid} and other stage-IV surveys.} The combination of \redit{spectroscopy from the Dark Energy Spectroscopic Instrument \citep[DESI;][]{desi2016}} with stage-III lensing data from UNIONS, and ultimately with stage-IV data from \textit{Euclid}, offers a unique path toward this goal.

Most direct IA measurements to date have used projected two-point correlation functions \citep{Mandelbaum2006,Joachimi2011,Singh2015,Samuroff2023}. \citet{Singh2023}\redit{, hereafter Si23,} recently demonstrated that a multipole-based estimator applied to the three-dimensional correlation function in redshift space provides significantly tighter constraints on IA model parameters\redit{ than projected estimators, by a factor of $1.5$--$2$ depending on the sample. The multipoles keep the full three-dimensional statistic, whereas $w_{g+}$ integrates it along the line of sight up to a fixed $\Pi_{\rm max}$ and loses information in the projection}. \redit{Three-dimensional, redshift-space treatments of alignments have also been developed by \citet{KuritaTakada2022} and \citet{Lamman2025}.}

In this work, we cross-correlate UNIONS galaxy shapes with DESI DR1 spectroscopic redshifts, \redit{as \citet{Siegel2025}, hereafter S26, did with four other imaging surveys, but using the multipole estimators of Si23.} \redit{We measure the clustering monopole and the IA quadrupole of the DESI Bright Galaxy Survey (BGS), luminous red galaxy (LRG) and emission-line galaxy (ELG) samples, split by galaxy colour and stellar mass, and fit them with the NLA model, and with TATT for the blue samples. Following \citet{Fortuna2025}, we then ask whether the alignment amplitude of red galaxies is set by halo mass: we measure the galaxy--galaxy lensing signal of UNIONS sources around the BGS red galaxies, fit it with a conditional stellar mass function, and use the resulting stellar-to-halo mass relation to assign an effective halo mass to every alignment sample. This tests the halo-mass scaling assumed by NLA-M and TATT-M down to the low-mass, faint end of the red population.} This paper is organised as follows. In Sect.~\ref{sec:data}, we describe the UNIONS and DESI DR1 datasets\redit{ and the alignment samples, and in Sect.~\ref{sec:measurements} the correlation-function, covariance and lensing measurements}. Section~\ref{sec:methodology} presents\redit{ the IA models and the likelihood}. We show our results \redit{for red and blue galaxies, including the halo-mass dependence of the alignment amplitude,} in Sect.~\ref{sec:results}\redit{, discuss them in Sect.~\ref{sec:discussion},} and conclude in Sect.~\ref{sec:conclusion}.

\section{Data}
\label{sec:data}

\begin{figure*}
   \centering
   \includegraphics[width=1.32\columnwidth]{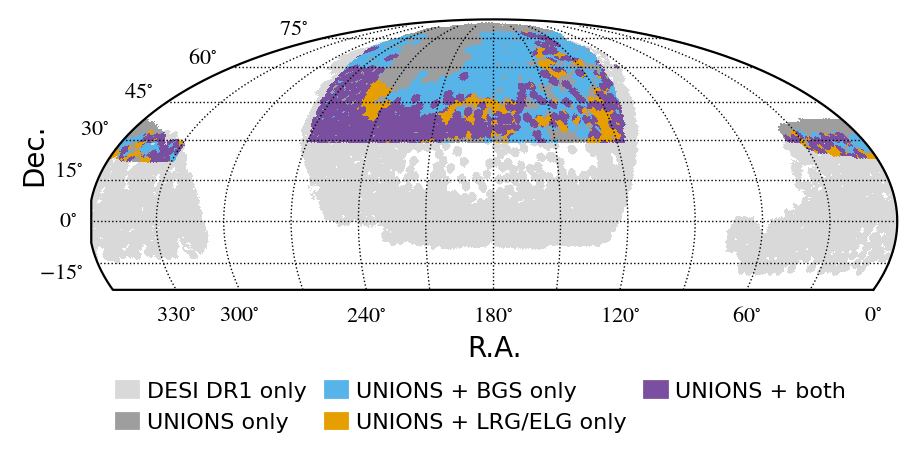}
   \caption{\redit{Sky coverage of DESI DR1 and of the UNIONS shape catalogue ($N_{\rm side}=128$). Coloured regions are covered by UNIONS and by the DESI bright-time programme only (BGS, blue), by the dark-time programme only (LRG and ELG, orange), or by both (purple); the BGS overlap is therefore the union of the blue and purple regions, and the LRG and ELG overlap that of the orange and purple regions.}}
   \label{fig:footprint}
\end{figure*}

The overlap between the UNIONS shape catalogue and DESI DR1 is \redit{shown} in Fig.~\ref{fig:footprint}\redit{: measured from the DESI random catalogues, the DESI DR1 samples overlap the UNIONS shape catalogue over $2397\,\mathrm{deg}^2$ (BGS), $1204\,\mathrm{deg}^2$ (LRG) and $1255\,\mathrm{deg}^2$ (ELG), the largest northern overlap of a stage-III lensing survey with DESI}.

\subsection{UNIONS shape catalogue}
\label{sec:unions}

\redit{The Ultraviolet Near-Infrared Optical Northern Survey \citep[UNIONS;][]{Gwyn2025} is a collaboration of wide-field imaging surveys} combining deep observations from the Canada-France-Hawaii Telescope (CFHT), Pan-STARRS, and Subaru in the Northern hemisphere. \redit{Its multi-band ($ugriz$) photometry serves a broad range of science, from weak lensing \citep{Robison2023,Li2024,Mpetha2025,Martin2026} and strong lensing \citep{Chan2022,Savary2022} to galaxy evolution \citep{Ellison2019,Bickley2021,Ferreira2024} and the Milky Way \citep{Jensen2021,Smith2024}, and supports photometric redshift estimation for the \textit{Euclid} mission.} In this work, we use galaxy shape measurements from the CFHT $r$-band data, obtained through the MegaPipe image processing pipeline \citep{Gwyn2008}.

\cedit{We use \redit{an extended version of the UNIONS shape catalogue of \citet{HervasPeters2026}\footnote{\redit{Internal \textsc{ShapePipe} v1.6.}}}.} Galaxy shapes are measured using the \redit{\textsc{ShapePipe}} pipeline \citep{Guinot2022,Farren2022}, which employs the \texttt{ngmix} software \citep{Sheldon2015} for model fitting and the \texttt{Metacalibration} framework \citep{SheldonHuff2017} for shear calibration. \texttt{Metacalibration} provides a self-calibrating approach to shear estimation by artificially shearing galaxy images and measuring the response of the shape estimator, yielding a shear response $\mathcal{R}$ that accounts for model bias, noise bias, and selection effects. Objects are detected and deblended with SExtractor \citep{Bertin96} and the point spread function (PSF) is modelled using \cedit{PSFEx \citep{Bertin2011}}.

\redit{Restricted to ${\rm Dec}\geq22^\circ$, the version of the shape catalogue we use covers $5400\,\mathrm{deg}^2$, of which $4800\,\mathrm{deg}^2$ lie in the Northern Galactic Cap and the rest in the Southern Galactic Cap,} with a median seeing of $0.69''$ in the $r$-band. \redit{This is $1900\,\mathrm{deg}^2$ more than the $3500\,\mathrm{deg}^2$ of the catalogue used for the UNIONS cosmic-shear analyses \citep{Guerrini2026,Goh2026} and for the previous UNIONS alignment measurement with BOSS/eBOSS \citep{HervasPeters2024}.} Each galaxy is assigned an inverse-variance shape weight \redit{$w_{\mathrm{SP}} = (2\sigma_{\mathrm{SN}}^2 + \sigma_{e_1}^2 + \sigma_{e_2}^2)^{-1}$}, where $\sigma_{\mathrm{SN}}$ is the galaxy shape noise and $\sigma_{e_i}$ is the \redit{standard deviation} of the ellipticity component $i = 1,2$ estimated by the model fitting procedure.\redit{\footnote{\redit{We have alternatively used the weights $w_{\rm des}$ adopted for the UNIONS cosmic-shear analyses \citep{HervasPeters2026,Guerrini2026,Goh2026} and found that they provide similar results.}}} For a detailed description of \redit{an earlier version (v1.4.6.3) of} the catalogue and its systematics validation, we refer to \cedit{\citet{HervasPeters2026} and \citet{Daley2026}}.

\subsection{DESI DR1 spectroscopic catalogues}
\label{sec:desi}

The Dark Energy Spectroscopic Instrument \citep{desi2016} is a multi-object fibre spectrograph mounted on the 4-metre Mayall telescope at Kitt Peak National Observatory, capable of simultaneously observing 5000 targets. The first data release \citep[DR1;][]{DESIDR1_2025} covers the first year of the main survey and provides spectroscopic redshifts for millions of galaxies and quasars over \redit{$9500\,\mathrm{deg}^2$ in each of the bright and dark programmes}.

We use the large-scale structure (LSS) catalogues constructed for the DESI DR1 cosmological analyses \citep[version 1.5;][]{DESI2024II,Ross2025}, which include matched random catalogues and \redit{the standard clustering weights. We weight every density sample with the clustering weight provided in these catalogues, $w_{\rm DESI}=w_{\rm comp}\,w_{\rm zfail}\,w_{\rm sys}$ which combines the completeness weight $w_{\rm comp}$, correcting for fibre-assignment incompleteness, the redshift-failure weight $w_{\rm zfail}$, and the imaging-systematics weight $w_{\rm sys}$ \citep[see][for details]{DESI2024II,Ross2025}.} The total weight assigned to each galaxy in the density sample is
\begin{equation}
   w_{\mathrm{tot}} = \redit{w_{\rm DESI}} \, w_{\mathrm{FKP}} \,,
   \label{eq:desi_weight}
\end{equation}
where $w_{\mathrm{FKP}}(z) = (1 + \bar{n}(z) P_0)^{-1}$ is the FKP weight \citep{FeldmanKaiserPeacock1994} with $P_0$ the power spectrum amplitude at $k \approx 0.15\,h\,\mathrm{Mpc}^{-1}$, which optimises the signal-to-noise of clustering measurements. \redit{The random catalogues carry the same weights as the data.} \cedit{\redit{DESI DR1 provides no imaging-systematics weights for BGS ($w_{\rm sys}=1$ in its LSS catalogues), so for BGS we recompute $w_{\rm DESI}$ in exactly the same way as \citet{Ross2025}, but with our own $w_{\rm sys}$, derived} following the same procedure as \citet{DESI2024II}: a linear regression of the galaxy density against the imaging-property maps\redit{. The BGS random catalogues are rebuilt accordingly. The correction modulates the density by $2\%$}, and we verified that removing it changes the galaxy--shape quadrupole by less than $0.1\sigma$ in every separation bin.} We consider the following tracer populations:

\noindent\textsf{Bright Galaxy Survey (BGS).} We use the BGS\_ANY sample \citep{Hahn2023}, which extends beyond the flux-limited BGS\_BRIGHT selection ($r < 19.5$) to include fainter targets observed during bright-time conditions.

\noindent\textsf{Luminous Red Galaxies (LRGs).} The LRG sample \citep{Zhou2023} is selected based on photometric colour and magnitude cuts designed to isolate massive, red, early-type galaxies at intermediate redshift, $0.4 < z < 1.1$. \redit{They} have been extensively studied in previous IA measurements \citep{hirata_intrinsic_2007,Joachimi2011,Singh2015,Samuroff2023}.

\noindent\textsf{Emission Line Galaxies (ELGs).} The ELG sample \citep{Raichoor2023} selects star-forming galaxies \redit{whose redshifts are measured from} their [O\,\textsc{ii}] emission\redit{; the DR1 clustering catalogues retain $0.8 < z < 1.6$ \citep{DESI2024II,Ross2025}}\cedit{; we use the \texttt{ELG\_LOPnotqso} LSS catalogue, i.e.\ the main (\texttt{LOP}) ELG selection with the objects also targeted as quasars removed}. These are blue, disc-dominated galaxies for which intrinsic alignments are expected to be weak. We use the ELG sample primarily as a null test for the IA signal.

\subsection{\texorpdfstring{\redit{Alignment samples}}{Alignment samples}}
\label{sec:samples}

To construct our analysis samples, we cross-match the DESI DR1 spectroscopic catalogues with the UNIONS shape catalogue. The DESI galaxies serve as the density (position) tracers, while the subset of DESI galaxies with UNIONS shape measurements constitutes the shape sample. We use a $k$-d tree matching in \texttt{astropy} with a tolerance of $1''$ to identify common objects, following the procedure of \citet{HervasPeters2024}. \cedit{No shape-quality selection is applied to this IA shape sample beyond the cross-match itself\redit{: the size, magnitude and signal-to-noise cuts of the lensing selection mainly serve to reject stars and spurious detections, which the match to spectroscopically confirmed galaxies already excludes}, and no PSF-leakage correction, which \citet{HervasPeters2024} showed to have a negligible impact on the IA statistics; the ellipticities of each sample are calibrated by its mean \texttt{Metacalibration} response $\mathcal{R}$ (Tables~\ref{tab:fits_red} and~\ref{tab:fits_blue}).}

\begin{figure}
   \centering
   \includegraphics[width=\columnwidth]{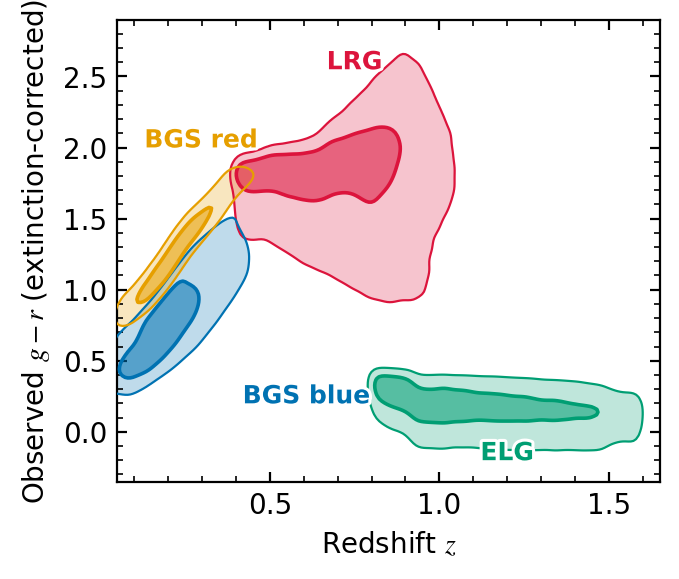}
   \caption{Observed \redit{extinction-corrected} $g-r$ colour as a function of redshift for the DESI DR1 galaxy samples used in this work. \redit{The filled regions enclose the $1\sigma$ (dark) and $2\sigma$ (light) levels of the density of each sample: BGS red (orange), BGS blue (blue), LRG (red) and ELG (green).}}
   \label{fig:colour_redshift}
\end{figure}

We further split the BGS sample by colour to study the colour dependence of the IA signal. Red and blue sub-samples are defined using a Gaussian mixture model (GMM) \redit{in the ($r$, $g-r$) magnitude--colour plane}. \cedit{Galaxies are binned in eight redshift slices with edges at $z=0.01$, $0.10$, $0.15$, $0.20$, $0.25$, $0.30$, $0.35$, $0.40$ and $0.50$ \redit{(the GMM is fitted on the full BGS catalogue, including the $z<0.1$ galaxies that are not used in the IA samples)}, and in each slice a three-component GMM is fitted. The components with the reddest and bluest mean colour define the red and blue sequences; their mean colours and widths are interpolated in redshift with a cubic spline\redit{, held constant beyond the first and last slice centres}, and each galaxy is assigned to the sequence it is closer to, in units of that sequence's width, provided it lies within $2\sigma$ of it. \redit{$1\%$} of the galaxies satisfy neither condition and are left out of both sub-samples.} The magnitudes used to perform this step are taken from the full DESI catalogue, and are corrected from galactic extinction. The resulting colour--redshift distribution of the BGS red, BGS blue, and LRG samples is shown in Fig.~\ref{fig:colour_redshift}. \redit{BGS red galaxies and LRGs are both red but are selected differently: the former are drawn from a flux-limited sample at $z<0.5$ by their position on the red sequence, the latter by optical--infrared colour cuts designed to yield an approximately constant comoving number density of massive galaxies \citep{Zhou2023}. The two samples overlap in stellar mass only above $\log(M_\star/h^{-2}M_\odot)=10.9$ (Table~\ref{tab:fits_red}).}

\cedit{Stellar masses are taken from the \textsc{FastSpecFit} value-added catalogue \citep{moustakas.etal23}, which fits the DESI spectra jointly with the Legacy Imaging Surveys photometry using stellar-population-synthesis templates at the measured spectroscopic redshift. We use its \texttt{LOGMSTAR} estimate throughout. \textsc{FastSpecFit} adopts $H_0 = 100\,{\rm km\,s^{-1}\,Mpc^{-1}}$ internally, so these masses are in units of $h^{-2}M_\odot$, and all stellar masses quoted in this work carry that convention.}

\section{\texorpdfstring{\redit{Measurements}}{Measurements}}
\label{sec:measurements}

\subsection{Correlation functions}
\label{sec:cf_estimation}

We estimate the three-dimensional correlation functions on two grids of pair separations: $(r_{\rm p},\Pi)$, where $r_{\rm p}$ is the transverse (projected) separation and $\Pi$ the line-of-sight separation; and $(s,\mu)$, where $s=\sqrt{r_{\rm p}^2+\Pi^2}$ is the three-dimensional separation and $\mu=\Pi/s$ is the cosine of the angle between the separation vector and the line of sight. \cedit{The clustering and intrinsic-alignment statistics are measured on two different logarithmic grids, both spanning $6$ to $100\,h^{-1}\,{\rm Mpc}$ in $s$ and in $r_{\rm p}$: a fine grid of $19$ bins for the clustering statistics ($\tilde\xi_{gg,0}$ and $w_{gg}$), and a coarser grid of $10$ bins for the shape cross-correlations ($\tilde\xi_{g+,2}$ and $w_{g+}$). In the $(r_{\rm p},\Pi)$ frame, $\Pi$ is binned into $20$ linear bins of width $10\,h^{-1}\,{\rm Mpc}$ over $\Pi\in[-100,100]\,h^{-1}\,{\rm Mpc}$; in the $(s,\mu)$ frame, $\mu$ is binned into $200$ linear bins of width $0.01$ over $\mu\in[-1,1]$.}

The galaxy--galaxy correlation function is estimated using the \citet{LandySzalay1993} estimator:
\begin{equation}
   \xi_{gg} = \frac{{\rm DD} - 2\,{\rm DR} + {\rm RR}}{{\rm RR}}\,,
   \label{eq:xi_gg_est}
\end{equation}
where ${\rm D}$ and ${\rm R}$ denote the data and random catalogues, respectively. All pair counts are normalised.

The galaxy--shape cross-correlation function is estimated with the estimator of \citet{Mandelbaum2006}:
\begin{equation}
   \xi_{g+} = \frac{{\rm S_+D} - {\rm S_+R_D}}{{\rm R_S R_D}}\,,
   \label{eq:xi_gp_est}
\end{equation}
where ${\rm S_+D} = \sum_{i\in{\rm S},j\in{\rm D}} w_i\,w_j\,e_+(i|j)$ is the weighted sum of the ellipticity component $e_+(i|j)$ of shape galaxy $i$ \redit{measured relative to the direction towards density galaxy $j$}, ${\rm S_+R_D}$ is the analogous sum using random points drawn from the density catalogue, and ${\rm R_S R_D}$ is the random--random normalisation.\cedit{\footnote{In practice we do not normalise by ${\rm R_S R_D}$ but by the density-random--shape pair count ${\rm R_D S}$, which is significantly faster to evaluate and has been shown to have a negligible impact on the measurement \citep{Johnston2019}.}} The subtraction of the shape--random term removes additive systematic biases arising from correlations between the shape measurement and the survey geometry. \redit{The component $e_+$ is positive when the major axis of the shape galaxy points towards the density galaxy (radial alignment) and negative when it is perpendicular to that direction (tangential alignment).}

\redit{The multipole pair counts are performed with the GPU-accelerated \textsc{cucount} code\footnote{\url{https://github.com/adematti/cucount}}, and the projected statistics with \textsc{pycorr} (galaxy--galaxy) and \textsc{TreeCorr} (galaxy--shape).} We use the DESI random catalogues, which match the angular and radial selection of each tracer sample\cedit{ (rebuilt for BGS as described in Sect.~\ref{sec:desi})}. For the density sample, we use the total weight $w_{\rm tot}$ (Eq.~\eqref{eq:desi_weight}). For the shape sample, each galaxy is weighted by the product of its shape weight $w_{\rm SP}$ (Sect.~\ref{sec:unions}) and the density completeness\cedit{, redshift-failure} and systematic weights $w_{\rm comp}\,\cedit{w_{\rm zfail}}\,w_{\rm sys}$, excluding the FKP weight.

\redit{Following Si23, we} decompose the anisotropic correlation functions into multipole moments. For a general correlation between fields $a$ and $b$ with spins $\mathfrak{s}_{ab}$, the multipole of order $\ell$ is
\begin{equation}
   \xi_{ab}^{\ell,\mathfrak{s}_{ab}}(r) = \frac{2\ell+1}{2}\,\frac{(\ell-\mathfrak{s}_{ab})!}{(\ell+\mathfrak{s}_{ab})!}\int\mathrm{d}\mu_r\;L^{\ell,\mathfrak{s}_{ab}}(\mu_r)\,\xi_{ab}(r,\mu_r)\,,
   \label{eq:multipole_def}
\end{equation}
where $\mu_r=\Pi/r$ is the cosine of the angle between the separation vector and the line of sight, $\Pi$ is the line-of-sight component, and $L^{\ell,\mathfrak{s}_{ab}}$ are associated Legendre polynomials. \redit{We denote the measured multipoles by $\tilde\xi_{ab,\ell}$, and use two of them: the clustering monopole $\tilde\xi_{gg,0}\equiv\xi^{0,0}_{gg}$ and the galaxy--shape quadrupole $\tilde\xi_{g+,2}\equiv\xi^{2,2}_{g+}$, both as functions of the redshift-space separation $s$. The projected correlation functions $w_{gg}(r_{\rm p})$ and $w_{g+}(r_{\rm p})$ are obtained by summing the $(r_{\rm p},\Pi)$ measurements over $|\Pi|<100\,h^{-1}\,{\rm Mpc}$.}

\cedit{We do not apply an additional transverse cut on $r_{\rm p}$ when building the multipoles. \redit{Si23} discarded pairs with $r_{\rm p}<5\,h^{-1}\,{\rm Mpc}$ from their multipole estimators, in order to suppress the contribution of redshift-space distortions along the line of sight. \redit{Applying the same cut to the low-redshift LRG sample changes the amplitude of $\tilde\xi_{g+,2}$ over the fitted range ($s>10\,h^{-1}\,{\rm Mpc}$) by $7\%$, i.e.\ $0.2\sigma$ of $A_{\rm IA}$, and its jackknife uncertainties by less than $3\%$}. We therefore retain all pairs.}

\subsection{\texorpdfstring{\redit{Covariance}}{Covariance}}
\label{sec:covariance}

\cedit{The covariance of all statistics is estimated by jackknife resampling: the overlapping survey area is divided into $N_{\rm jk}=70$ equal-area regions by running the $K$-means algorithm on the random catalogue of each tracer, and $N_{\rm jk}$ correlation functions are computed, each with one region removed. \redit{This number is set so that a BGS region, $6^\circ$ across, spans the largest fitted separation, $100\,h^{-1}\,{\rm Mpc}$, at the effective redshift of the BGS density sample; the LRG and ELG regions are larger, as these samples lie at higher redshift.} The resulting covariance is corrected for the cross-region pairs (pairs whose two members lie in different regions) following \citet{MohammadPercival2022}. We do not use the cross-covariance between $\tilde\xi_{gg,0}$ and $\tilde\xi_{g+,2}$ in the likelihood, as the limited number of jackknife regions makes the off-diagonal blocks noisy\redit{; Si23 noted that neglecting it should not have a large impact, since these covariances are dominated by noise}.}

\subsection{\texorpdfstring{\redit{Galaxy--galaxy lensing}}{Galaxy--galaxy lensing}}
\label{sec:ggl}

\cedit{\redit{To assign halo masses to the alignment samples (Sect.~\ref{sec:results_mass_evolution}), we measure the excess surface mass density $\Delta\Sigma(r_{\rm p})$ around the BGS red lens samples with the \texttt{dsigma} package \citep{dsigma}. The sources are UNIONS galaxies \citep{HervasPeters2026} selected in photometric redshift well behind the $z<0.5$ lenses, and their redshift distribution is calibrated with clustering redshifts against the DESI spectroscopic samples \citep{dAssignies2025}. The source selection, the estimator and the calibration are described in Appendix~\ref{app:shmr}, together with the stellar-to-halo mass relation derived from this measurement.}}

\section{\texorpdfstring{\redit{Modelling and inference}}{Modelling and inference}}
\label{sec:methodology}

\redit{Our baseline is the non-linear linear alignment (NLA) model \citep{Catelan2001,HirataSeljak2004,Bridle07}, in which the intrinsic shape field responds linearly to the large-scale tidal tensor $s_{ij}$, computed from the non-linear rather than the linear matter power spectrum, with a single free amplitude $A_{\rm IA}$, complemented by a linear galaxy bias $b_1$ and the linear-order Kaiser prescription for redshift-space distortions \citep{Kaiser1987}. In our convention a positive $A_{\rm IA}$ corresponds to shapes elongated along the stretching direction of the tidal field, i.e.\ pointing towards overdensities (radial alignment), as expected for pressure-supported elliptical galaxies, and a negative $A_{\rm IA}$ to shapes perpendicular to it (tangential alignment). For the blue samples, for which the tidal-torquing term is expected to be most relevant, we also fit the tidal alignment and tidal torque (TATT) model \citep{Blazek19}, which expands the intrinsic shape field to second order in $s_{ij}$ and in the matter overdensity $\delta$:}
\begin{equation}
   \gamma^{\rm I}_{ij} = \redit{C_1}\,s_{ij} + \redit{C_{1\delta}}\,\delta\,s_{ij} + \redit{C_2}\left(s_{ik}s_{kj} - \tfrac{1}{3}\,\delta_{ij}\,s^2\right),
   \label{eq:tatt_field}
\end{equation}
where $\redit{C_1}$ describes the linear response to the tidal field (tidal alignment), $\redit{C_{1\delta}}$ the modulation of that response by the local galaxy density, and $\redit{C_2}$ the quadratic tidal-torquing contribution\redit{, which arises when the angular momentum of disc galaxies is acquired from tidal torques and their shapes follow their spin}. \redit{Their relation to the amplitudes $A_{\rm IA}$, $b_{\rm TA}$ and $A_2$ quoted in this paper is given in Appendix~\ref{app:tatt_coeffs}.} \redit{The NLA model retains only the linear term ($b_{\rm TA}=A_2=0$). In the TATT fits we set $b_{\rm TA}=1$, keeping $b_1$, $A_{\rm IA}$ and $A_2$ free. The corresponding real- and redshift-space power spectra are given in Appendix~\ref{app:ia_model}.} \cedit{In the following we consider only the linear bias approximation: the higher-order terms arising from non-linear galaxy bias in the power spectra are not included \citep{Blazek19,Paviot2026}.}

\subsection{Model prediction}
\label{sec:multipoles}

The model predictions for each multipole \redit{(Eq.~\eqref{eq:multipole_def})} are obtained through Hankel transforms of the redshift-space power spectra \cedit{(Eqs.~\eqref{eq:kaiser} and \eqref{eq:pgs_rsd} in Appendix~\ref{app:ia_model})}:
\begin{equation}
   \xi_{ab}^{\ell,\mathfrak{s}_{ab}}(r) = \frac{i^\ell}{2\pi^2}\int \mathrm{d}k\;k^2\,\alpha_{\ell,ab}\,P_{ab}(k)\,j_\ell(kr)\,,
   \label{eq:hankel}
\end{equation}
where $j_\ell$ are spherical Bessel functions and the coefficients $\alpha_{\ell,ab}$ encode the redshift-space distortion contributions. \cedit{The projected correlation functions follow from the same decomposition by integrating the multipole expansion along the line of sight (Eq.~19 of \redit{Si23}):}
\begin{equation}
   w_{ab}(r_{\rm p}) = \sum_\ell 2\int_0^{\Pi_{\rm max}}\mathrm{d}\Pi\;\xi_{ab}^{\ell,\mathfrak{s}_{ab}}(r)\,L^{\ell,\mathfrak{s}_{ab}}\!\left(\frac{\Pi}{r}\right),
   \label{eq:projected}
\end{equation}
where $r=\sqrt{r_{\rm p}^2+\Pi^2}$\redit{, which accounts} for the residual effects of redshift-space distortions at finite $\Pi_{\rm max}$. \cedit{For galaxy clustering ($\mathfrak{s}_{ab}=0$), the non-vanishing multipoles are the monopole ($\ell=0$), quadrupole ($\ell=2$), and hexadecapole ($\ell=4$); for the galaxy--shape cross-correlation ($\mathfrak{s}_{ab}=2$), they are the spin-2 quadrupole ($\ell=2$) and hexadecapole ($\ell=4$). The corresponding coefficients $\alpha_{\ell,ab}$ are given in Appendix~\ref{app:rsd} (Eqs.~\eqref{eq:a0gg}--\eqref{eq:a4gp}).} \redit{We evaluate all correlation functions at point separations (no bin averaging) and use $\Pi_{\rm max}=100\,h^{-1}\,{\rm Mpc}$, as in the measurements.} \cedit{We neglect the gravitational-lensing contamination of the galaxy--shape correlation, i.e.\ the shear of the shape galaxies by the matter around the density galaxies and the magnification of either sample \citep[the $gG$, $\mu I$ and $\mu G$ terms of][]{Samuroff2023}: adding these terms to the model, with the magnification coefficients measured for our samples \redit{with the method of \citet{Wenzl2024} (Appendix~\ref{app:shmr})}, shifts $A_{\rm IA}$ by less than $0.1\sigma$ in every sample, because both samples are spectroscopic and the contamination only becomes appreciable at the largest separations, where the quadrupole carries little weight.}

\subsection{\texorpdfstring{\redit{Fitted statistics and scale cuts}}{Fitted statistics and scale cuts}}
\label{sec:scale_cuts}
\label{sec:statistics}
\redit{We fit the clustering monopole $\tilde\xi_{gg,0}$ and the galaxy--shape quadrupole $\tilde\xi_{g+,2}$ jointly, and, as a cross-check, the projected $w_{gg}$ and $w_{g+}$ (Eq.~\eqref{eq:projected}). We adopt conservative scale cuts, summarised in Table~\ref{tab:scale_cuts}, to exclude the small scales where non-linear clustering and Fingers-of-God effects are not captured by our linear-order model. The clustering monopole, which is the more sensitive to these effects, is cut at $30\,h^{-1}\,{\rm Mpc}$ on its finer grid (Sect.~\ref{sec:cf_estimation}), well inside the linear regime, and the shape quadrupole at $10\,h^{-1}\,{\rm Mpc}$. For the TATT fits the quadrupole cut is relaxed to $6\,h^{-1}\,{\rm Mpc}$, since the tidal-torquing term contributes predominantly on small scales. The projected cross-check uses the cut commonly adopted in previous analyses, $r_{\rm p}>6\,h^{-1}\,{\rm Mpc}$ for both $w_{gg}$ and $w_{g+}$ \citep{Joachimi2011,Singh2015,Johnston2019}, which admits pairs down to a three-dimensional separation $s=6\,h^{-1}\,{\rm Mpc}$ (since $s\geq r_{\rm p}$); Si23 fitted their multipoles from $s=5\,h^{-1}\,{\rm Mpc}$. Our baseline multipole cuts are therefore more conservative than both.}

\begin{table}
\caption{\redit{Scale cuts of the fits. Separations are $s$ for the multipoles and $r_{\rm p}$ for the projected statistics; all statistics are fitted up to $100\,h^{-1}\,{\rm Mpc}$. $N_{\rm pts}$ is the number of fitted points.}}
\label{tab:scale_cuts}
\centering
\begin{tabular}{llcc}
\hline\hline
Statistic & Model & $r_{\rm min}$ [$h^{-1}\,{\rm Mpc}$] & $N_{\rm pts}$ \\
\hline
$\tilde\xi_{gg,0}$ & NLA, TATT & $30$ & $8$ \\
$\tilde\xi_{g+,2}$ & NLA & $10$ & $8$ \\
$\tilde\xi_{g+,2}$ & TATT & $6$ & $10$ \\
$w_{gg}$ & NLA & $6$ & $19$ \\
$w_{g+}$ & NLA & $6$ & $10$ \\
\hline
\end{tabular}
\end{table}

\subsection{Likelihood and parameter inference}
\label{sec:likelihood}

The cosmology is fixed to the \cedit{\citet{Planck2018}} best-fit values ($h=0.6766$, $\Omega_ch^2=0.11933$, $\Omega_bh^2=0.02242$, \redit{$A_s=2.105\times10^{-9}$}, $n_s=0.9665$).

We assume a Gaussian likelihood
\begin{equation}
   \chi^2 = \sum_{\alpha\in\{0,2\}}\left(\boldsymbol{d}_\alpha - \boldsymbol{m}_\alpha\right)^T\mathbf{C}_\alpha^{-1}\left(\boldsymbol{d}_\alpha - \boldsymbol{m}_\alpha\right),
   \label{eq:chi2}
\end{equation}
where $\boldsymbol{d}_\alpha$ and $\boldsymbol{m}_\alpha$ are the data and model vectors for the monopole ($\alpha=0$) and quadrupole ($\alpha=2$), respectively, and $\mathbf{C}_\alpha$ is the corresponding jackknife covariance block \redit{(Sect.~\ref{sec:covariance})}. \cedit{We do not apply the \citet{Hartlap2007} correction to the inverse covariance\redit{, which is derived for covariances estimated from independent realisations, such as mocks, and does not strictly apply to jackknife resamples of a single data set.}}

\cedit{The IA model, the likelihood and the fits are implemented in the public \textsc{ia2pt} package\footnote{\url{https://github.com/rpaviot/IA2pt} (version 0.2.0); documentation at \url{https://ia2pt.readthedocs.io}.}.} We sample the posterior with the \textsc{nautilus} nested sampler \citep{nautilus}, using flat priors $b_1\in[0.1,4]$ and $A_{\rm IA}\in[-10,10]$\cedit{, and $A_2\in[-6,6]$ for the TATT fits}. The sampler is run until the effective sample size reaches $n_{\rm eff}=50\,000$. \redit{Quoted uncertainties are $1\sigma$ marginalised posterior widths, and significances are the posterior median in units of this width.} The theoretical predictions are computed using \textsc{pyccl} \citep{pyccl} with \redit{\textsc{CAMB} \citep{Lewis2000}} and \textsc{HMCode2020} \citep{Mead2021} for the non-linear matter power spectrum, and \textsc{FAST-PT} \redit{\citep{McEwen2016,Fang2017}} for the perturbation-theory integrals. Hankel transforms from Fourier to configuration space are performed using the routines in \textsc{FAST-PT}. \cedit{The clustering monopole is evaluated at the effective redshift of the density sample, and the galaxy--shape quadrupole (and $w_{g+}$) at the pair-weighted effective redshift of the density--shape cross-correlation, $z_{\rm IA}$, listed in the tables of Appendix~\ref{app:corrfuncs}. \redit{The same $b_1$ is used at both redshifts, without evolution. Evolving it from the density to the pair redshift at constant $b_1D(z)$ instead changes $A_{\rm IA}$ by at most $6\%$ ($0.6\sigma$, for the full BGS red sample; median shift $0.08\sigma$) and leaves the $\chi^2$ unchanged.}}

\section{Results}
\label{sec:results}

\redit{All fits use the NLA model applied to $\tilde\xi_{gg,0}$ and $\tilde\xi_{g+,2}$, except the TATT fits of Sect.~\ref{sec:results_tatt}; we interpret the results in Sect.~\ref{sec:discussion}.}

\subsection{Red galaxies}
\label{sec:results_red}

\begin{figure}
   \centering
   \includegraphics[width=\columnwidth]{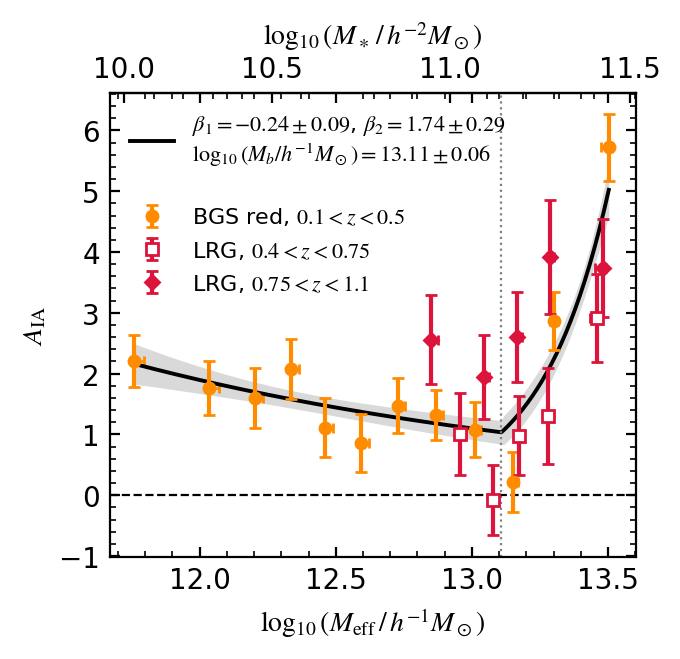}
   \caption{\redit{Alignment amplitude of the red samples as a function of the effective halo mass of each bin (bottom axis) and of its median stellar mass (top axis), the two being related by the relation of Appendix~\ref{app:shmr}. Orange circles show the BGS red bins, open squares and filled diamonds the low- and high-redshift LRG bins. Horizontal bars show the $16$--$84\%$ interval of $M_{\rm eff}$, not included in the fits. The black line and grey band show the best-fit broken power law (Eq.~\eqref{eq:broken_pl}) and its $1\sigma$ uncertainty, and the dotted line its pivot mass.}}
   \label{fig:AIA_red}
\end{figure}

\redit{As expected, red galaxies are radially aligned. Their best-fit parameters are collected in Table~\ref{tab:fits_red}, and their amplitudes are shown as functions of stellar mass and halo mass in Fig.~\ref{fig:AIA_red}. The correlation functions of the BGS red mass bins are shown in Fig.~\ref{fig:bgs_snr_xi2} (Appendix~\ref{app:corrfuncs}).}

\subsubsection{BGS red}
\label{sec:results_bgs_red}

\cedit{\redit{The density sample, common to the BGS red and blue fits, is the full BGS clustering catalogue, whose linear bias is $b_1 = 1.39 \pm 0.04$.}}

\cedit{We split the BGS red sample into twelve stellar-mass bins \redit{built to yield comparable expected signal-to-noise in $\tilde\xi_{g+,2}$, which makes them equipopulated at low mass and progressively sparser at high mass\footnote{\redit{To predict the signal-to-noise, the binning assumes an amplitude that is constant below $\log(M_\star/h^{-2}M_\odot)=11$ and grows as $M_\star^{0.5}$ above it, following the bright-end luminosity dependence of \citet{Fortuna21}.}}}.
\redit{The mass dependence is non-monotonic (Fig.~\ref{fig:AIA_red}): the amplitude decreases from the lowest-mass bins to a dip at $10.6\leq\log(M_\star/h^{-2}M_\odot)\leq11.2$, and rises steeply in the two highest-mass bins, up to $A_{\rm IA}=5.7\pm0.6$ at $\log(M_\star/h^{-2}M_\odot)=11.4$.}}

\subsubsection{LRG}
\label{sec:results_lrg}

\cedit{The LRG sample is split into two redshift bins: a low-redshift bin at $0.4 < z < 0.75$ and a high-redshift bin at $0.75 < z < 1.1$.}

\cedit{\redit{We split each LRG redshift bin into five stellar-mass bins, in a similar way as for the BGS sample. The} amplitude increases with mass in both redshift bins, \redit{up to $A_{\rm IA}=3$--$4$ in the most massive bin ($\log(M_\star/h^{-2}M_\odot) = 11.4$). At the same median stellar mass, the high-redshift LRGs are more strongly aligned than the low-redshift ones: the global amplitudes ($A_{\rm IA}=1.17\pm0.35$ and $2.95\pm0.38$) differ at $3.4\sigma$, and every high-redshift mass bin lies above its low-redshift counterpart. The low-redshift bin at $\log(M_\star/h^{-2}M_\odot) = 11.1$ is consistent with zero, while the low-redshift LRG bins otherwise agree with the BGS red bins of similar mass (Fig.~\ref{fig:AIA_red}). We discuss this redshift difference in Sect.~\ref{sec:disc_lrg}.}}

\subsubsection{\texorpdfstring{\redit{Halo-mass dependence}}{Halo-mass dependence}}
\label{sec:results_mass_evolution}

\redit{Figure~\ref{fig:AIA_red} also shows the amplitudes of the BGS red and LRG bins against the effective halo mass $M_{\rm eff}$ of each bin.} \cedit{Following \citet{Fortuna2025}, $M_{\rm eff}$ is the mean mass of the haloes hosting a central galaxy,
\begin{equation}
   M_{\rm eff}(z) = \frac{\int \langle N_{\rm c}|M,z\rangle\, n(M,z)\, M\, {\rm d}M}{\int \langle N_{\rm c}|M,z\rangle\, n(M,z)\, {\rm d}M}\,,
   \label{eq:meff}
\end{equation}
where $n(M,z)$ is the halo mass function at redshift $z$ and $\langle N_{\rm c}|M,z\rangle$ the central occupation of the best-fit conditional stellar mass function (Appendix~\ref{app:shmr})\redit{, integrated over the stellar-mass range of the bin}. Satellites therefore do not enter $M_{\rm eff}$; halo masses are defined at 200 times the mean matter density and quoted in $h^{-1}M_\odot$. The relation between median stellar mass and $M_{\rm eff}$ is established on the volume-limited BGS red lens bins and applied to the alignment samples through their median stellar mass; for the LRG samples this is an extrapolation in both stellar mass and redshift.}

\redit{We fit power laws to the measured $A_{\rm IA}$ and their uncertainties with \textsc{iminuit} \citep{iminuit}. A single power law,
\begin{equation}
   A_{\rm IA} = A_0\left(\frac{M_{\rm eff}}{10^{13}\,h^{-1}M_\odot}\right)^{\beta_{\rm M}},
   \label{eq:single_pl}
\end{equation}
the form adopted by \citet{Fortuna2025}, provides a poor description of our measurements over the full mass range, with $\chi^2=96.8$ for 20 degrees of freedom (hereafter $\chi^2/{\rm dof}=96.8/20$). Restricting it to $M_{\rm eff}>10^{12}\,h^{-1}M_\odot$ gives $\beta_{\rm M}=0.66\pm0.18$, but the fit remains poor ($\chi^2/{\rm dof}=80.2/19$), mostly because the three lowest-mass BGS red bins above $10^{12}\,h^{-1}M_\odot$ lie well above the power law and the low-redshift LRG bin that is consistent with zero lies well below it. A broken power law,
\begin{equation}
   A_{\rm IA} = A_b\left(\frac{M_{\rm eff}}{M_b}\right)^{\beta},\qquad
   \beta=\left\{\begin{array}{ll}\beta_1\,, & M_{\rm eff}<M_b\,,\\ \beta_2\,, & M_{\rm eff}\geq M_b\,,\end{array}\right.
   \label{eq:broken_pl}
\end{equation}
continuous at a free pivot mass $M_b$, improves the fit substantially ($\chi^2/{\rm dof}=32.3/18$), with a nearly flat faint end ($\beta_1=-0.24\pm0.09$) below $M_b=1.3\times10^{13}\,h^{-1}M_\odot$ and a steep rise above it ($\beta_2=1.74\pm0.29$; Fig.~\ref{fig:AIA_red}).}

The low-mass behaviour does not depend on the extrapolation to the LRGs. Restricting the fits to the BGS red samples alone, for which the mapping is direct\redit{\footnote{\redit{We estimate the $M_{\rm eff}$ of the IA samples from the SHMR determined on the volume-limited samples, by interpolating between the lens bins; evaluating Eq.~\eqref{eq:meff} directly over each IA sample changes it by at most $0.03$\,dex and the broken power law by less than $0.5\sigma$.}}}, the single power law remains strongly disfavoured (\redit{$\chi^2/{\rm dof}=69.5/10$}) while the broken power law provides a good description of the data (\redit{$\chi^2/{\rm dof}=8.2/8$}), \redit{with the same faint-end slope and pivot as the joint fit}.

\subsection{Blue galaxies}
\label{sec:results_blue}

\begin{figure}
   \centering
   \includegraphics[width=\columnwidth]{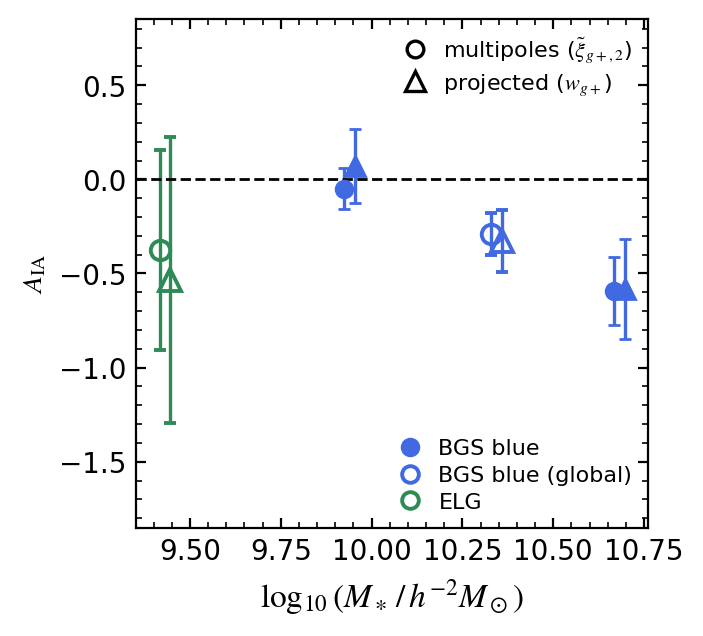}
   \caption{Alignment amplitude as a function of median stellar mass for \redit{the BGS blue (blue) and ELG (green) samples. Filled symbols are the stellar-mass bins, open symbols the global samples; circles are multipole and triangles projected fits.}}
   \label{fig:AIA_blue}
\end{figure}

\begin{figure}
   \centering
   \includegraphics[width=\columnwidth]{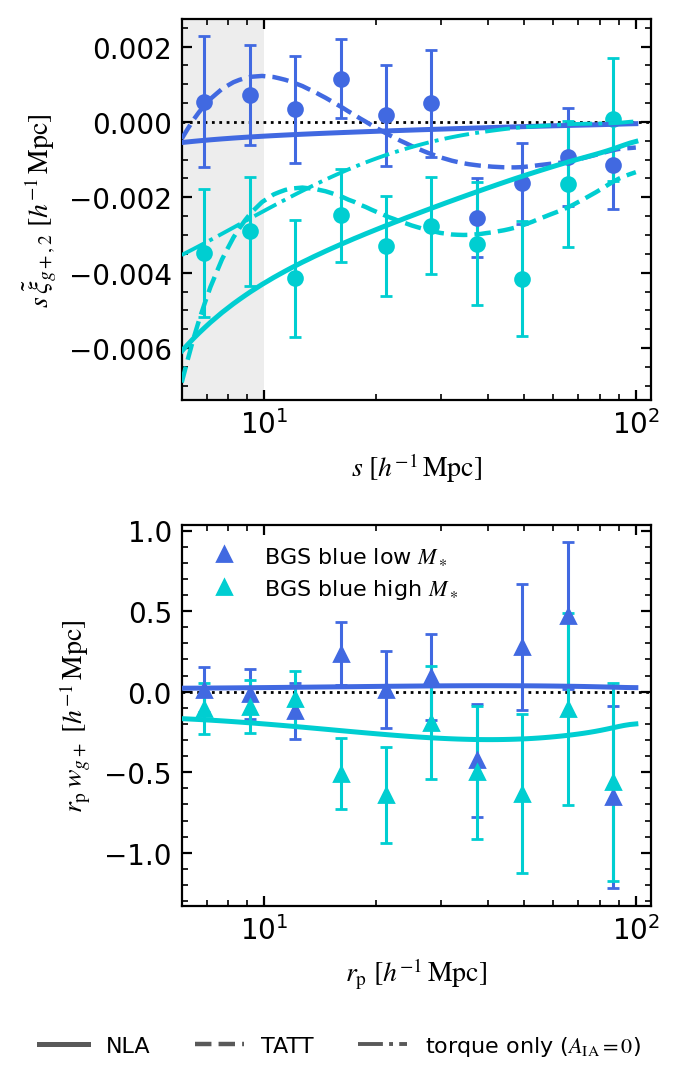}
   \caption{\redit{Galaxy--shape correlation functions of the low- (dark blue) and high-mass (cyan) BGS blue bins. \textit{Top:} $s\,\tilde\xi_{g+,2}$ with the best-fit NLA (solid) and TATT (dashed) models, and the torque-only model ($A_{\rm IA}=0$, dash-dotted) for the high-mass bin. \textit{Bottom:} $r_{\rm p}\,w_{g+}$ with the best-fit NLA model. The grey band marks the scales excluded from the NLA quadrupole fit but used by the TATT and torque-only fits. Negative values correspond to tangential alignment.}}
   \label{fig:corrfunc_blue}
\end{figure}

\redit{Unlike the red galaxies, the blue samples show no radial alignment, and the massive BGS blue galaxies are tangentially aligned. The NLA fits are collected in Table~\ref{tab:fits_blue}, the amplitudes are shown as a function of stellar mass in Fig.~\ref{fig:AIA_blue}, and the correlation functions of the BGS blue mass bins in Fig.~\ref{fig:corrfunc_blue}.} \cedit{\redit{The multipole and projected ($w_{g+}$) fits} agree in every sample, and the multipoles are consistently the more precise: the uncertainty on $A_{\rm IA}$ from $w_{g+}$ is \redit{$43$--$78\%$} larger than from the multipoles on the same data (last column of Table~\ref{tab:fits_blue}).}

\begin{table*}
\caption{Best-fit NLA model parameters for the \emph{blue} galaxy samples (BGS blue and ELG) from the multipole analysis ($\tilde\xi_{gg,0}$ and $\tilde\xi_{g+,2}$) and from the projected statistic $w_{g+}$. The corresponding TATT constraints are given in Table~\ref{tab:fits_tatt}.}
\label{tab:fits_blue}
\centering
\small
\setlength{\tabcolsep}{3pt}
\begin{tabular}{lcccccccccc}
\hline\hline
Sample & $z$ range & $z_{\rm IA}$ & $\log M_\star$ & $N_{\rm shapes}$ & $\sigma_e$ & $\mathcal{R}$ & $b_1$ & $A_{\rm IA}^{\tilde\xi_{g+,2}}\ (\chi^2_\nu)$ & $A_{\rm IA}^{w_{g+}}\ (\chi^2_\nu)$ & Gain \\
\hline
\multicolumn{11}{c}{\textit{BGS blue}} \\
\hline
Global & $0.1$--$0.5$ & $0.25$ & $10.34 \pm 0.53$ & $827\,884$ & $0.226$ & $0.92$ & $1.390 \pm 0.040$ & $-0.291 \pm 0.113\ (1.80)$ & $-0.328 \pm 0.166\ (1.21)$ & $+47\%$ \\
$M_\star$ bin 1 (low) & $0.1$--$0.5$ & $0.20$ & $9.94 \pm 0.33$ & $405\,662$ & $0.232$ & $0.93$ & -- & $-0.049 \pm 0.111\ (1.89)$ & $0.069 \pm 0.197\ (1.40)$ & $+78\%$ \\
$M_\star$ bin 2 (high) & $0.1$--$0.5$ & $0.30$ & $10.68 \pm 0.23$ & $405\,663$ & $0.218$ & $0.90$ & -- & $-0.592 \pm 0.180\ (1.52)$ & $-0.585 \pm 0.266\ (1.15)$ & $+48\%$ \\
\hline
\multicolumn{11}{c}{\textit{ELG}} \\
\hline
Global & $0.8$--$1.6$ & $1.11$ & $9.43 \pm 0.65$ & $414\,759$ & $0.299$ & $0.72$ & $1.215 \pm 0.042$ & $-0.375 \pm 0.533\ (1.76)$ & $-0.535 \pm 0.760\ (1.30)$ & $+43\%$ \\
\hline
\end{tabular}
\tablefoot{$z_{\rm IA}$ as in Table~\ref{tab:fits_red}; both $\tilde\xi_{g+,2}$ and $w_{g+}$ are modelled at $z_{\rm IA}$, the monopole and $w_{\rm gg}$ at the density-sample redshift ($0.35$ (BGS blue), $1.18$ (ELG)). The last column is the gain in precision on $A_{\rm IA}$ delivered by the multipole analysis relative to the projected one, $\sigma_{w_{g+}}/\sigma_{\tilde\xi_{g+,2}}-1$.}
\end{table*}

\subsubsection{BGS blue}
\label{sec:results_bgs_blue}

\cedit{For the full BGS blue sample we measure $A_{\rm IA} = -0.29 \pm 0.11$ from the multipole analysis, \redit{a mild tangential alignment at $2.6\sigma$, consistent with the projected statistic ($A_{\rm IA} = -0.33 \pm 0.17$)}.}

\cedit{We split the BGS blue sample into two stellar-mass bins. The low-mass bin is \redit{consistent with zero ($A_{\rm IA} = -0.05 \pm 0.11$)}. The high-mass bin \redit{is tangentially aligned, with $A_{\rm IA} = -0.59 \pm 0.18$ ($3.3\sigma$), which the projected statistic confirms at lower significance ($2.2\sigma$); to our knowledge, this is the most significant detection of a non-zero alignment of blue galaxies to date}. The mild tangential signal of the global sample is thus driven entirely by the high-mass half of the blue population.} \redit{The detection does not rely on the NLA model. Without any model, the eight fitted quadrupole points of the high-mass bin are inconsistent with zero at $2.2\sigma$ ($\chi^2=16.9$, $p=0.03$). The amplitude is also stable against the quadrupole scale cut: for $r_{\rm min}$ between $6$ and $20\,h^{-1}\,{\rm Mpc}$ it stays tangential and within the uncertainty of the fiducial value (Fig.~\ref{fig:rmin_sweep}).}

\cedit{\redit{The reduced $\chi^2$ of the multipole fits (Table~\ref{tab:fits_blue}) are high, but the excess comes from the BGS blue monopole that the three samples share rather than from the alignment signal (Appendix~\ref{app:corrfuncs}).}}

\subsubsection{ELG}
\label{sec:results_elg}

\cedit{The ELG sample at $0.8 < z < 1.6$ yields $A_{\rm IA} = -0.37 \pm 0.53$ from the multipole analysis\redit{, consistent with zero and with the projected statistic, and constrains the ELG alignment amplitude to $-1.4 < A_{\rm IA} < 0.7$ at $2\sigma$. The multipole fit is somewhat poor ($\chi^2_\nu=1.76$), with the excess shared equally between the monopole and the quadrupole (Appendix~\ref{app:corrfuncs}). The ELG correlation functions are shown in Fig.~\ref{fig:corrfunc_elg} (Appendix~\ref{app:corrfuncs}).}}

\subsubsection{\cedit{TATT fits}}
\label{sec:results_tatt}

\begin{figure}
   \centering
   \includegraphics[width=\columnwidth]{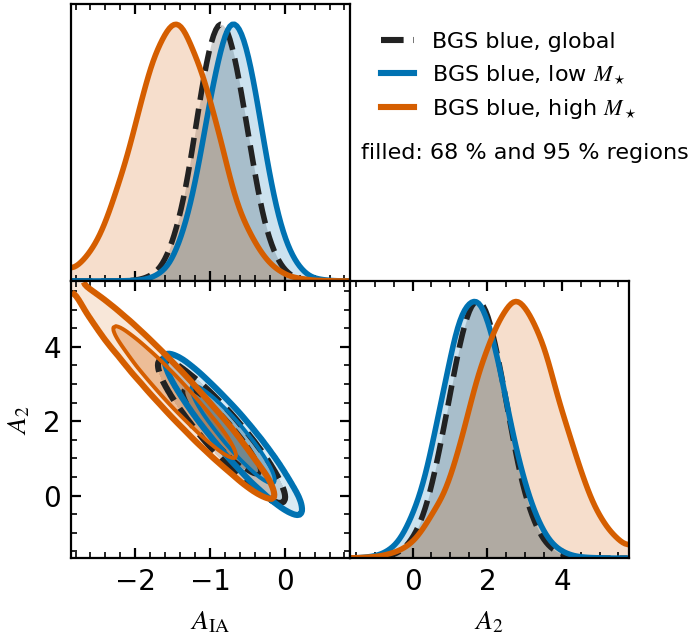}
   \caption{\redit{Posteriors of the TATT amplitudes $A_{\rm IA}$ (tidal alignment) and $A_2$ (tidal torquing) for the three BGS blue samples, with the $68\%$ and $95\%$ credible regions and the marginal distributions. The ELG posterior, which fills the $A_2$ prior, is omitted.}}
   \label{fig:tatt_contours}
\end{figure}

\cedit{\redit{To test whether the tangential signal requires a tidal-torquing contribution, we repeat the NLA multipole fits of the blue samples with the full TATT model} (Sect.~\ref{sec:methodology}), relaxing the quadrupole scale cut to $r_{\rm min}=6\,h^{-1}\,{\rm Mpc}$ (\redit{Table~\ref{tab:scale_cuts}}). For the global BGS blue sample we obtain \redit{$A_{\rm IA} = -0.85 \pm 0.33$ and $A_2 = 1.73 \pm 0.75$, and the same pattern, a negative $A_{\rm IA}$ and a positive $A_2$, holds in both stellar-mass bins (Table~\ref{tab:fits_tatt}). The galaxy bias is unchanged from the NLA fits.}}

\cedit{These marginal constraints should not be read as two independent measurements. \redit{The two amplitudes are almost perfectly anti-correlated ($\rho = -0.99$; Fig.~\ref{fig:tatt_contours}), because with the normalisation of Eq.~\eqref{eq:tatt_coeffs} a positive $A_2$ produces a radial $\tilde\xi_{g+,2}$, like a positive $A_{\rm IA}$: around a spherical overdensity, the quadratic term aligns the shapes with the direction of largest tidal eigenvalue in absolute value, which is the radial one. The well-measured quantity is therefore $A_{\rm IA}+c\,A_2$, with $c=0.4$ from the posteriors, which is determined six to eight times more precisely than $A_{\rm IA}$ alone. This combination reproduces the NLA results: null at low mass ($0.00\pm0.05$) and tangential at high mass ($-0.19\pm0.07$).

In our fits the tangential signal is carried by the negative tidal-alignment amplitude, while the positive $A_2$ partly cancels it on small scales; the positive marginal $A_2$ is therefore not evidence for a tidal-torquing origin of the tangential alignment. Fitted to the same scales ($r_{\rm min}=6\,h^{-1}\,{\rm Mpc}$), TATT lowers the $\chi^2$ of the NLA model by only $2.8$--$3.9$ for one additional parameter, and the Bayesian evidence does not favour it ($\ln B_{\rm TATT/NLA}=-0.4$ to $-0.7$): the data do not require the tidal-torquing term. Conversely, a torque-only fit ($A_{\rm IA}=0$, $A_2$ free) recovers the tangential signal of the high-mass bin as a negative $A_2=-0.38\pm0.15$, but describes it less well than NLA on the same scales ($\Delta\chi^2=4.6$ for the same number of parameters, $\ln B_{\rm TT/NLA}=-1.6$), because the quadratic term falls off faster with separation than the measured signal. Setting $b_{\rm TA}=0$ instead of unity leaves this picture unchanged: the high-mass sample remains tangentially aligned at a similar significance, the marginal $A_2$ drops by a factor of 2--4, and the evidence favours NLA slightly more ($\ln B_{\rm TATT/NLA}=-1$).} The ELG TATT fit is unconstrained: \redit{the $A_2$ posterior piles up against the lower edge of its prior ($A_2=-6$), so Table~\ref{tab:fits_tatt} quotes only an upper limit on $A_2$, and the $A_{\rm IA}$ posterior, which follows $A_2$ along the degeneracy, reflects the prior range rather than the data. Its fit is also poor ($\chi^2_\nu = 2.21$; Appendix~\ref{app:corrfuncs}).}}

\section{\texorpdfstring{\redit{Discussion}}{Discussion}}
\label{sec:discussion}

\subsection{\texorpdfstring{\redit{Halo-mass dependence of red-galaxy alignments}}{Halo-mass dependence of red-galaxy alignments}}
\label{sec:disc_mass}

\redit{\citet{Fortuna2025} found that a single power law with $\beta_{\rm M}=0.44\pm0.04$ describes the halo-mass dependence of the alignment amplitude over $1.9\times10^{12}$--$3.7\times10^{14}\,h^{-1}M_\odot$, the range of the compilation they fitted, their own KiDS LRG samples spanning $2.7\times10^{12}$--$2.6\times10^{13}\,h^{-1}M_\odot$, and \citet{Herle2026} reached the same conclusion in the FLAMINGO simulations above $10^{12}\,h^{-1}M_\odot$, with $\beta_{\rm M}=0.77\pm0.015$. Above $10^{12}\,h^{-1}M_\odot$ our single-power-law slope, $\beta_{\rm M}=0.66\pm0.18$, is consistent with both values, but the single power law is a poor fit, and the broken power law is strongly preferred (Sect.~\ref{sec:results_mass_evolution}). Our measurements do not extend to the cluster masses ($>10^{14}\,h^{-1}M_\odot$) that anchor the high-mass end of their sample. The flattening is not confined to masses below those probed by \citet{Fortuna2025}: the break and our BGS red bins at $12.4\leq\log(M_{\rm eff}/h^{-1}M_\odot)\leq12.9$ lie within the range of their KiDS LRG samples, whose amplitudes there ($A_{\rm IA}=1$--$4$, with large uncertainties) lie mostly above ours ($A_{\rm IA}=0.9$--$1.5$). Within this common mass range, the difference may reflect the photometric versus spectroscopic selections or the different definitions of the red samples.}

\cedit{Halo mass may also not be the only variable at play: using the FLAMINGO simulations, \citet{Herle2026b} showed that the IA amplitude also depends on halo formation redshift, with earlier-forming haloes being more strongly aligned. If the faint red BGS galaxies preferentially occupied early-formed haloes, their alignment would be enhanced relative to the mass-only expectation and the $A_{\rm IA}$--$M_{\rm eff}$ relation would flatten at the faint end. \redit{In FLAMINGO, however, the formation time mainly shifts the amplitude while leaving its slope with mass unchanged, so this mechanism flattens the relation only if the mix of formation times changes with mass, early-forming haloes being increasingly represented at low mass. Testing the formation-time interpretation requires an observational proxy, such as the S\'ersic index, the specific star formation rate or the stellar age \citep{Herle2026b}; we leave it to future work.}}

\subsection{\texorpdfstring{\redit{Redshift dependence of LRG alignments}}{Redshift dependence of LRG alignments}}
\label{sec:disc_lrg}

\redit{The high-redshift LRGs are more strongly aligned than the low-redshift ones at the same stellar mass (Sect.~\ref{sec:results_lrg}), and we have not identified the origin of this difference. Since the alignment amplitude of red galaxies increases with luminosity \citep[][S26]{Fortuna21}, a difference could arise if the high-redshift LRGs, selected from a flux-limited sample, were brighter at fixed stellar mass. They are, but only slightly: in common $0.1$\,dex stellar-mass slices, their \textsc{FastSpecFit} rest-frame $r$-band absolute magnitudes (k-corrected to $z=0.1$) are brighter by $0.1$--$0.24$\,mag, as expected from passive evolution. With the bright-end luminosity slope $\beta=1.2$ of \citet{Fortuna21}, this raises the expected amplitude by at most $30\%$, far short of the factor of $2.5$ between the two global amplitudes, so luminosity at fixed stellar mass does not explain the difference. The orientation-dependent target selection of DESI LRGs \citep{Lamman2024} and stellar-mass systematics at $z=0.9$ could also contribute.}

\subsection{\texorpdfstring{\redit{Tangential alignment of massive blue galaxies}}{Tangential alignment of massive blue galaxies}}
\label{sec:disc_blue}

\redit{The tangential alignment of the massive BGS blue galaxies is three times weaker than the radial alignment of the BGS red sample ($|A_{\rm IA}|=0.6$ against $1.7$). The massive blue galaxies are as strongly star-forming as the low-mass ones: their median specific star formation rate from \textsc{FastSpecFit}, $\log({\rm sSFR}/{\rm yr}^{-1})=-11.07$, is the same as that of the low-mass bin ($-11.03$) and $2.7$\,dex above that of the BGS red sample ($-13.80$). In the linear alignment picture, a tangential alignment requires a negative linear response of the shapes to the tidal field, for which no standard mechanism exists: tidal stretching of pressure-supported galaxies yields a radial alignment. Tidal torquing, in which disc galaxies acquire their angular momentum from the tidal field and their shapes follow their spin, is the natural candidate. Within TATT, \citet{Blazek19} associated a tangential alignment of spiral galaxies with a negative $A_2$, the opposite of the sign we recover, but also showed that tidal torquing generates, on large scales, a contribution with the same scale dependence as the linear tidal alignment. A tangential alignment produced by torquing could therefore be absorbed by a negative $A_{\rm IA}$ in our fits, without any signature in $A_2$. Since TATT does not improve on NLA, and a torque-only model fits the signal less well than a linear one (Sect.~\ref{sec:results_tatt}), the data favour the scale dependence of a negative linear alignment, but our measurements cannot determine whether this linear term is itself the large-scale imprint of tidal torquing.}

\redit{Hydrodynamical simulations do not offer a consensus either. In Horizon-AGN, disc (blue) galaxies are oriented tangentially around spheroidal (red) ones in three dimensions, an effect that is suppressed in projection \citep{Chisari15}. Comparing TNG300, Horizon-AGN and EAGLE with common shape and morphology definitions, \citet{vanHeukelum2026} found that disc galaxies align radially or not at all; the tangential alignment survives only in Horizon-AGN, for kinematically selected discs around ellipticals at $z=1$ with reduced shapes, which down-weight the galaxy outskirts. In IllustrisTNG \citep{TNG2018}, by contrast, the shapes of massive disc galaxies align with the tidal field \citep{Shi2021}, and blue and spiral galaxies show a radial linear alignment \citep{Samuroff21,Zjupa2022}. A tidal-torquing alignment appears only for the most massive spirals at $z=1$ in TNG100 \citep{Zjupa2022}, and between $z=0$ and $z=1$ in MillenniumTNG \citep{Pakmor2023,Delgado2023}.} \redit{The sign of the alignment of disc galaxies in simulations thus depends on how the discs are selected, by colour or by kinematics, on how their shapes are measured, and on their stellar mass \citep{vanHeukelum2026}. Our blue samples are selected by colour, and their shapes are measured by fitting a single Gaussian profile, which, like the reduced shapes, gives most weight to the inner parts of the galaxy, so a quantitative comparison requires simulated galaxies selected and measured in the same way, at the stellar masses of our high-mass bin. Whatever its origin, a non-zero alignment of massive blue galaxies confirms that restricting cosmic-shear sources to blue galaxies reduces, but does not remove, the IA contamination, so that such analyses must still marginalise over it \citep{McCullough2026}.}

\subsection{\texorpdfstring{\redit{Comparison with S26}}{Comparison with S26}}
\label{sec:disc_s25}

\redit{S26 measured IA with the same DESI DR1 tracers cross-matched with four imaging surveys (DES, KiDS, HSC, SDSS), using the projected statistic $w_{g+}$. For red galaxies the two analyses agree: both detect IA at $>10\sigma$ for the BGS red sample and in the LRGs. S26 described their amplitudes by luminosity and colour alone, with no residual stellar-mass dependence, but the dip of our amplitude at intermediate stellar mass is also visible in the stellar-mass panel of their Fig.~8. For blue galaxies, neither analysis finds a radial alignment ($A_{\rm IA}=0.1\pm0.3$ for their bluest BGS galaxies), and their bright blue BGS galaxies show a tangential alignment, $A_{\rm IA}=-0.4\pm0.2$ (read from their figure), close to our high-mass value. The blue amplitudes cannot be compared one to one, since they depend strongly on where the colour boundary is drawn: relaxing the S26 cut from $M_r-M_z<0.5$ to $0.65$ raises their amplitude to $1.4\pm0.2$. For ELGs, S26 reported $A_{\rm IA}=-3.0\pm1.0$ at $1.15<z<1.55$ and an amplitude consistent with zero below; our constraint over the combined range sits between the two. Their TATT fits, from $2\,h^{-1}\,{\rm Mpc}$ with $b_{\rm TA}$ and a second-order bias free, gave $A_2$ consistent with zero, which our $A_2$, not constrained independently of $A_{\rm IA}$, does not contradict.}

\section{Conclusions}
\label{sec:conclusion}

\cedit{We have measured the intrinsic alignments of DESI DR1 galaxies by
cross-correlating the BGS, LRG, and ELG spectroscopic samples with the UNIONS
weak-lensing shape catalogue, over the \redit{area common to the two surveys,
$2400\,{\rm deg}^2$ for BGS and $1200$--$1250\,{\rm deg}^2$ for the LRG and
ELG samples}. \redit{We fitted} the clustering monopole $\tilde\xi_{gg,0}$ and the
galaxy--shape quadrupole $\tilde\xi_{g+,2}$ \redit{of the three-dimensional
correlation function} jointly with the NLA model. Each
\redit{red} sample was assigned an effective halo mass through a conditional stellar mass
function fitted to the galaxy--galaxy lensing signal measured around \redit{the BGS
red galaxies}, so that the alignment amplitude could be followed as a function of halo
mass rather than of a luminosity or stellar-mass proxy alone.}

\redit{Red galaxies are strongly and radially aligned. In stellar-mass
bins, the amplitude dips at intermediate masses and rises steeply at the
massive end. As a function of effective halo mass, it is not described by a
single power law over $6\times10^{11}$--$3\times10^{13}\,h^{-1}M_\odot$, even
when the fit is restricted to $M_{\rm eff}>10^{12}\,h^{-1}M_\odot$. A
broken power law is strongly preferred, with a nearly flat faint
end below $M_b=1.3\times10^{13}\,h^{-1}M_\odot$ and a steep rise above
it. The same break is recovered from the BGS red samples alone, for which the
stellar-to-halo mass relation is measured rather than extrapolated. The
flattening lies largely within the mass range of the KiDS LRG samples of
\citet{Fortuna2025}, so it is a tension with their single power law within
the common mass range, possibly due to the different selections of the red samples.}

\redit{Blue galaxies are not aligned in the same sense. The mild tangential
alignment of the full BGS blue sample comes entirely from its high-mass half,
which shows a tangential alignment at $3.3\sigma$ ($2.2\sigma$ without any
model of its scale dependence), while the low-mass half is consistent with zero.
We also fitted TATT to the blue samples. Although the tidal-alignment and
tidal-torquing amplitudes are strongly degenerate, the tangential signal is
carried by a negative tidal-alignment amplitude, while the tidal-torquing
amplitude comes out slightly positive, i.e.\ of the sign that produces a radial
alignment. A torque-only fit describes the data less well than NLA, and the
Bayesian evidence does not favour TATT over NLA. The tangential alignment is
therefore not produced by the quadratic tidal-torquing term of TATT, although
tidal torquing could still contribute through the effective linear term it
generates on large scales \citep{Blazek19}. The ELG sample yields no detection and
constrains the alignment amplitude to $-1.4<A_{\rm IA}<0.7$ at $2\sigma$ over
$0.8<z<1.6$.}

\redit{On identical data, the multipole estimators recover the amplitudes of $w_{g+}$ with
uncertainties smaller by $20$--$43\%$ (Appendix~\ref{app:wgp_multipoles}),
less than the factor of $1.5$--$2$ reported by Si23, in part because of our more
conservative scale cuts.}

\redit{By reaching the faint end of the red population and blue galaxies down
to $\log(M_\star/h^{-2}M_\odot)=10$, these measurements extend direct IA
constraints towards the populations that dominate cosmic-shear source
samples.} \cedit{These results carry a direct implication for cosmological analyses. IA
priors are usually built by extrapolating a luminosity- or mass-dependent power
law calibrated on bright, massive samples down to the faint populations that
dominate cosmic-shear source catalogues. \redit{Our measurements indicate that
a single power law does not capture the alignment of low-mass red galaxies:
below the break the amplitude is nearly flat, at $A_{\rm IA}=1$--$2$, so
that a single power law fitted across the break mispredicts it in both
directions, too high just below the break and too low at the lowest masses.}
Mass-dependent IA models of the NLA-M or TATT-M family
\citep{NLA-M,Herle2026} should allow for a break rather than a single slope.
That the relation flattens rather than continues to fall may also point to a
second variable, such as halo formation time \citep{Herle2026b}, which our data
cannot isolate.}

\cedit{\redit{Three} limitations should temper these conclusions. \redit{First, the halo-mass scale of the $A_{\rm IA}$--$M_{\rm eff}$ relation rests on the abundance anchor of the stellar-to-halo mass relation. That anchor depends on how the fibre-assignment incompleteness of BGS is corrected: with individual-inverse-probability weights, the effective halo masses shift coherently, while the shape of the relation and the break are unchanged (Appendix~\ref{app:shmr}). A fibre-complete BGS sample is therefore needed to pin the mass scale.} \redit{Second, the galaxy bias is modelled as linear and constant, with no scale dependence and no evolution between the density and pair redshifts. This is safe for the clustering monopole, fitted above $30\,h^{-1}\,{\rm Mpc}$, but the shape quadrupole, in which $b_1$ and $A_{\rm IA}$ are degenerate, extends down to $10\,h^{-1}\,{\rm Mpc}$ ($6\,h^{-1}\,{\rm Mpc}$ for TATT), where a scale-dependent bias would be absorbed into the alignment amplitudes.} \redit{Finally, NLA is an empirical description, so the
amplitudes quoted here are effective amplitudes for the fitted scales. This
matters most for the TATT fits of the blue samples, which rely on scales down to
$6\,h^{-1}\,{\rm Mpc}$, where perturbative IA models describe the alignment
signal poorly; the TATT amplitudes and the model comparison should therefore be
regarded as indicative rather than established.} DESI DR2, combined with the full UNIONS footprint, will enlarge
both the overlapping area and the samples, raise the fibre-assignment
completeness of BGS through its additional passes, and allow the measurements
to be repeated on large scales alone: it will make it possible to \redit{confirm the tangential alignment
of massive blue galaxies and to test whether the tidal-torquing signal survives on large scales}, to split the ELG sample in redshift, and to
calibrate the halo masses of the LRG lenses directly instead of by
extrapolation.}

\begin{acknowledgements}
RP, AC and FHP acknowledge financial support from the Centre national d'\'etudes
spatiales (CNES), France (ROR: \url{https://ror.org/04h1h0y33}).
MJH and LVW acknowledge support from NSERC through Discovery Grants.
GSHP acknowledges support from the Pan-STARRS project.
We are honoured and grateful for the opportunity of observing the Universe from
Maunakea and Haleakala, which both have cultural, historical and natural
significance in Hawai'i. This work is based on data obtained as part of the
Canada-France Imaging Survey, a CFHT large program of the National Research
Council of Canada and the French Centre National de la Recherche Scientifique.
Based on observations obtained with MegaPrime/MegaCam, a joint project of CFHT
and CEA Saclay, at the Canada-France-Hawai'i Telescope (CFHT) which is operated
by the National Research Council (NRC) of Canada, the Institut National des
Sciences de l'Univers (INSU) of the Centre National de la Recherche Scientifique
(CNRS) of France, and the University of Hawai'i. This research used the
facilities of the Canadian Astronomy Data Centre operated by the National
Research Council of Canada with the support of the Canadian Space Agency. This
research is based in part on data collected at Subaru Telescope, which is
operated by the National Astronomical Observatory of Japan. Pan-STARRS is a
project of the Institute for Astronomy of the University of Hawai'i, and is
supported by the NASA SSO Near Earth Observation Program under grants
80NSSC18K0971, NNX14AM74G, NNX12AR65G, NNX13AQ47G, NNX08AR22G, 80NSSC21K1572
and by the State of Hawai'i. This work was made possible by utilising the
CANDIDE cluster at the Institut d'Astrophysique de Paris. The cluster was
funded through grants from the PNCG, CNES, DIM-ACAV, the Euclid Consortium, and
the Danish National Research Foundation Cosmic Dawn Center (DNRF140). It is
maintained by St\'ephane Rouberol. We gratefully acknowledge support from the
CNRS/IN2P3 Computing Center (Lyon - France) for providing computing and
data-processing resources needed for this work. The authors acknowledge the use
of the Canadian Advanced Network for Astronomy Research (CANFAR) Science
Platform operated by the Canadian Astronomy Data Centre (CADC) and the Digital
Research Alliance of Canada (DRAC), with support from the National Research
Council of Canada (NRC), the Canadian Space Agency (CSA), CANARIE, and the
Canada Foundation for Innovation (CFI). LWKG thanks the University of Edinburgh
School of Physics and Astronomy for a postdoctoral Fellowship.
We also used \textsc{numpy} \citep{Harris2020}, \textsc{scipy}
\citep{Virtanen2020}, \textsc{matplotlib} \citep{Hunter2007}, \textsc{TreeCorr}
\citep{Jarvis2004,TreeCorr2015}, \textsc{pycorr} with \textsc{Corrfunc}
\citep{corrfunc}, \textsc{healpy} \citep{Zonca2019}, \textsc{HEALPix}
\citep{Gorski2005}, \textsc{skyproj}, \textsc{astropy} \citep{astropy2013,
astropy2018, astropy2022}, \textsc{iminuit} \citep{iminuit}, and
\textsc{scikit-learn} \citep{Pedregosa2011}.
The analysis is specified and executed with \textsc{lightcone-cli} (\url{https://github.com/LightconeResearch/lightcone-cli}) following the
ASTRA specification; the complete pipeline is available at \url{https://github.com/rpaviot/unions-desi-ia-paper}.
The analysis pipeline, the reproducibility directory and the manuscript were
developed and edited with the assistance of Claude (Anthropic), used through
Claude Code; the author has verified all results and takes full responsibility
for the content of this paper.
\end{acknowledgements}

\bibliographystyle{bibtex/aa}
\bibliography{bibtex/references}

\begin{appendix}
\section{\texorpdfstring{\redit{Intrinsic-alignment model and estimators}}{Intrinsic-alignment model and estimators}}
\label{app:ia_model}

\subsection{\texorpdfstring{\redit{Alignment model}}{Alignment model}}
\label{app:tatt_coeffs}
\label{app:nla}

\redit{Following the \textsc{pyccl} implementation, the coefficients of the TATT expansion (Eq.~\eqref{eq:tatt_field}) are related to the dimensionless amplitudes quoted throughout this paper by
\begin{equation}
   C_1 = -A_{\rm IA}\,\frac{\bar{C}_1\rho_{\rm crit}\,\Omega_{\rm m}}{D(z)}\,,\quad
   C_{1\delta} = b_{\rm TA}\,C_1\,,\quad
   C_2 = 5\,A_2\,\frac{\bar{C}_1\rho_{\rm crit}\,\Omega_{\rm m}}{D^2(z)}\,,
   \label{eq:tatt_coeffs}
\end{equation}
where $\bar{C}_1=5\times10^{-14}\,h^{-2}M_\odot^{-1}\,{\rm Mpc}^3$ \citep{Bridle07}, $\rho_{\rm crit}$ is the critical density at $z=0$, so that $\bar{C}_1\rho_{\rm crit}=0.0139$ is a dimensionless constant, and $D(z)$ is the linear growth factor normalised to unity at $z=0$; all the redshift and cosmology dependence is carried by $\Omega_{\rm m}/D(z)$.}

\redit{The NLA model \citep{Catelan2001,HirataSeljak2004,Bridle07} retains only the linear term of the TATT expansion, with $C_1$ given by Eq.~\eqref{eq:tatt_coeffs}.} \cedit{The matter--intrinsic power spectrum in the NLA model, and the galaxy--intrinsic cross-power spectrum in real space that follows from linear galaxy bias, read}
\begin{equation}
   \cedit{P_{g{\rm I}}(k,z) = b_1\,P_{\delta{\rm I}}(k,z) = -A_{\rm IA}\,b_1\,\redit{\bar{C}_1}\rho_{\rm crit}\,\frac{\Omega_{\rm m}}{\redit{D(z)}}\,P_\delta(k,z)\,,}
   \label{eq:PdeltaI}
\end{equation}
where $P_\delta$ is the non-linear matter power spectrum\cedit{, which we compute with \textsc{HMCode2020}\redit{, using its \textsc{CAMB} implementation as wrapped in \textsc{pyccl}}. The same non-linear $P_\delta$ enters the clustering spectrum $P_{gg}$ (Eq.~\eqref{eq:kaiser}) and the galaxy--intrinsic spectrum $P_{g{\rm I}}$}. Galaxy shapes are spin-2 quantities; we restrict our analysis to $E$-modes throughout.

\cedit{The higher-order terms of the TATT expansion (Eq.~\eqref{eq:tatt_field}) generate additional contributions to $P_{g{\rm I}}$ involving one-loop perturbation-theory integrals. We do not reproduce them here: the full expressions for the tidal-alignment, density-weighting and tidal-torquing contributions to the intrinsic-alignment power spectra are given in \citet{Blazek19}, whose notation and conventions we follow, and we evaluate them with \textsc{FAST-PT} as implemented in \textsc{pyccl}.}

\begin{figure}
   \centering
   \includegraphics[width=\columnwidth]{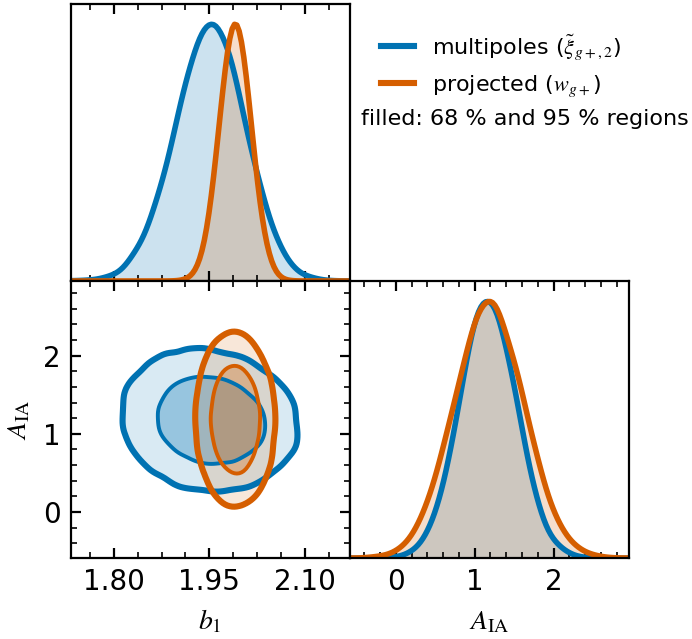}
   \caption{Posterior distributions of $b_1$ and $A_{\rm IA}$ for the low-redshift
   LRG sample, from the multipoles \redit{($\tilde\xi_{gg,0}$ and $\tilde\xi_{g+,2}$, blue)} and the projected $w_{g+}$
   statistic (orange). The
   \redit{filled} contours enclose 68\% and 95\% of the posterior.}
   \label{fig:gtc_wgp_multipoles}
\end{figure}

\subsection{Redshift-space distortions}
\label{app:rsd}

Peculiar velocities along the line of sight induce anisotropies in the estimated three-dimensional separations, known as redshift-space distortions (RSDs). At linear order, the Kaiser effect \citep{Kaiser1987} modifies the power spectra by factors of $(1+\beta\mu_k^2)$, where $\beta=f/b_1$ is the ratio of the linear growth rate $f$ to the linear galaxy bias $b_1$, and $\mu_k$ is the cosine of the angle between $\boldsymbol{k}$ and the line of sight. \cedit{The galaxy clustering and galaxy--shape cross-power spectra in redshift space read \redit{(Si23)}}
\begin{align}
   \cedit{P_{gg}^s(k,\mu_k,z)} &\cedit{= b_1^2\left[1+\beta\mu_k^2\right]^2 P_\delta(k,z)\,,} \label{eq:kaiser}\\
   \cedit{P_{g+}^s(k,\mu_k,z)} &\cedit{= \redit{\left(1-\mu_k^2\right)}\left(1+\beta\mu_k^2\right) P_{g{\rm I}}(k,z)\,,} \label{eq:pgs_rsd}
\end{align}
where $P_{g{\rm I}}$ is the real-space galaxy--intrinsic power spectrum defined in Eq.~\eqref{eq:PdeltaI}. These angular-dependent factors give rise to quadrupole and hexadecapole contributions, which we model via the multipole formalism \cedit{of Sect.~\ref{sec:multipoles}. They enter the Hankel transforms of Eq.~\eqref{eq:hankel} through the coefficients $\alpha_{\ell,ab}$. For galaxy clustering ($\mathfrak{s}_{ab}=0$), these read \citep{Baldauf2010,Singh2021}}
\begin{align}
   \alpha_{0,gg} &= 1 + \tfrac{2}{3}\beta + \tfrac{1}{5}\beta^2\,, \label{eq:a0gg}\\
   \alpha_{2,gg} &= \tfrac{4}{3}\beta + \tfrac{4}{7}\beta^2\,, \label{eq:a2gg}\\
   \alpha_{4,gg} &= \tfrac{8}{35}\beta^2\,, \label{eq:a4gg}
\end{align}
\cedit{and for the galaxy--shape cross-correlation ($\mathfrak{s}_{ab}=2$),}
\begin{align}
   \alpha_{2,g+} &= \tfrac{1}{3}\left(1+\tfrac{1}{7}\beta\right), \label{eq:a2gp}\\
   \alpha_{4,g+} &= \tfrac{2}{105}\beta\,. \label{eq:a4gp}
\end{align}

\subsection{Consistency of the projected and multipole estimators}
\label{app:wgp_multipoles}

Our baseline constraints are derived from the multipoles
$\tilde\xi_{gg,0}$ and $\tilde\xi_{g+,2}$, while most IA studies in the literature use
the projected statistic $w_{g+}$. Figure~\ref{fig:gtc_wgp_multipoles}
compares the posterior distributions of the galaxy bias $b_1$ and the IA
amplitude $A_{\rm IA}$ obtained from the two estimators for the low-redshift LRG
sample ($0.4 < z < 0.75$). \cedit{The two posteriors are in
excellent agreement, with $A_{\rm IA} = 1.17 \pm 0.35$ from the multipoles and
$A_{\rm IA} = 1.18 \pm 0.44$ from $w_{g+}$: the two estimators recover the same
amplitude, \redit{and the multipole uncertainty is $20\%$ smaller}. The gain is larger for the
blue samples, \redit{with $w_{g+}$ uncertainties larger by factors of $1.43$ to $1.78$}
(last column of Table~\ref{tab:fits_blue}), but remains below the factor of \redit{$1.5$--$2$}
reported by \redit{Si23}. \redit{Part of this difference comes from the scale cuts. Si23 removed pairs with $r_{\rm p}<5\,h^{-1}\,{\rm Mpc}$ (Sect.~\ref{sec:multipoles}) and fitted both multipoles down to $5\,h^{-1}\,{\rm Mpc}$, whereas we fit $\tilde\xi_{gg,0}$ from $30$ and $\tilde\xi_{g+,2}$ from $10\,h^{-1}\,{\rm Mpc}$ (Sect.~\ref{sec:scale_cuts}), while our projected statistics use $r_{{\rm p},\min}=6\,h^{-1}\,{\rm Mpc}$ for}
both $w_{gg}$ and $w_{g+}$. The multipole fits thus rely on fewer and larger
scales than the projected ones, so the improvement quoted here should be
regarded as a lower bound on what the estimator can deliver.}

\FloatBarrier 

\section{\texorpdfstring{\redit{Fit results and correlation functions}}{Fit results and correlation functions}}
\label{app:corrfuncs}

\cedit{This appendix gathers the best-fit parameters and the correlation-function measurements underlying the fits of Sect.~\ref{sec:results}\redit{ that are not shown in the main text}.}

\redit{The $\chi^2$ of the BGS blue multipole fits (Table~\ref{tab:fits_blue}) is driven by the clustering monopole that the three samples share ($\chi^2=15.3$ for eight points); the alignment quadrupole gives $\chi^2=11.2$, $6.1$ and $10.0$ for eight points in the low-mass, high-mass and global samples. For the ELG, the monopole and the quadrupole contribute about equally ($\chi^2=11.9$ and $12.8$, eight points each). The ELG TATT fit ($\chi^2_\nu=2.21$) has no single outlier: its largest quadrupole residual ($2.5\sigma$) lies at $s=21\,h^{-1}\,{\rm Mpc}$, well inside the NLA range.}

\begin{figure}[!hb]
   \centering
   \includegraphics[width=\columnwidth]{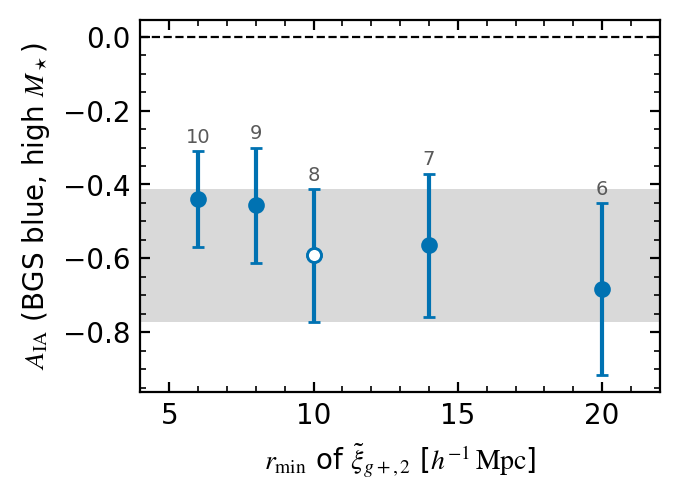}
   \caption{\redit{NLA alignment amplitude of the high-mass BGS blue bin as a function of the minimum separation $r_{\rm min}$ of the fitted quadrupole. The open circle and grey band show the fiducial fit; the numbers give the fitted quadrupole points. The fits share most of their data, so the points are strongly correlated.}}
   \label{fig:rmin_sweep}
\end{figure}

\begin{table}
\caption{Best-fit TATT model parameters for the blue samples from the multipole analysis. The density-weighting bias is fixed to $b_{\rm TA}=1$, leaving $b_1$, $A_{\rm IA}$ and $A_2$ free; the fits retain linear galaxy bias only.}
\label{tab:fits_tatt}
\centering
\small
\begin{tabular}{lcccc}
\hline\hline
Sample & $z_{\rm IA}$ & $A_{\rm IA}$ & $A_2$ & $\chi^2_\nu$ \\
\hline
BGS blue, global & $0.25$ & $-0.85 \pm 0.33$ & $1.73 \pm 0.75$ & $1.58$ \\
BGS blue, low $M_\star$ & $0.20$ & $-0.69 \pm 0.36$ & $1.64 \pm 0.85$ & $1.53$ \\
BGS blue, high $M_\star$ & $0.30$ & $-1.46 \pm 0.54$ & $2.76 \pm 1.16$ & $1.39$ \\
ELG, global & $1.11$ & $-0.07 \pm 0.87$ & $<0.96$ & $2.21$ \\
\hline
\end{tabular}
\tablefoot{The sample properties ($z$ range, $\log M_\star$, $N_{\rm shapes}$, $\sigma_e$, $\mathcal{R}$) are listed in Table~\ref{tab:fits_blue} and are not repeated here; the recovered $b_1$ agrees with the NLA value of Table~\ref{tab:fits_blue} to better than $0.5\%$ in every case. $A_{\rm IA}$ is the tidal-alignment amplitude (the same parameter as in the NLA fits of Table~\ref{tab:fits_blue}) and $A_2$ the tidal-torquing amplitude, normalised as in Eq.~\eqref{eq:tatt_coeffs}. The quadrupole scale cut is relaxed to $r_{\rm min}=6\,h^{-1}\,{\rm Mpc}$ for these fits, retaining all ten $\tilde\xi_{g+,2}$ points, while the monopole cut is unchanged at $30\,h^{-1}\,{\rm Mpc}$; these constraints therefore rely on smaller scales than the NLA baseline, so $\chi^2_\nu$ is not directly comparable with the NLA values of Table~\ref{tab:fits_blue}. The ELG $A_2$ posterior piles up against the lower edge of its prior ($-6<A_2<6$), so we quote its $95\%$ upper limit.}
\end{table}

\begin{figure}
   \centering
   \includegraphics[width=\columnwidth]{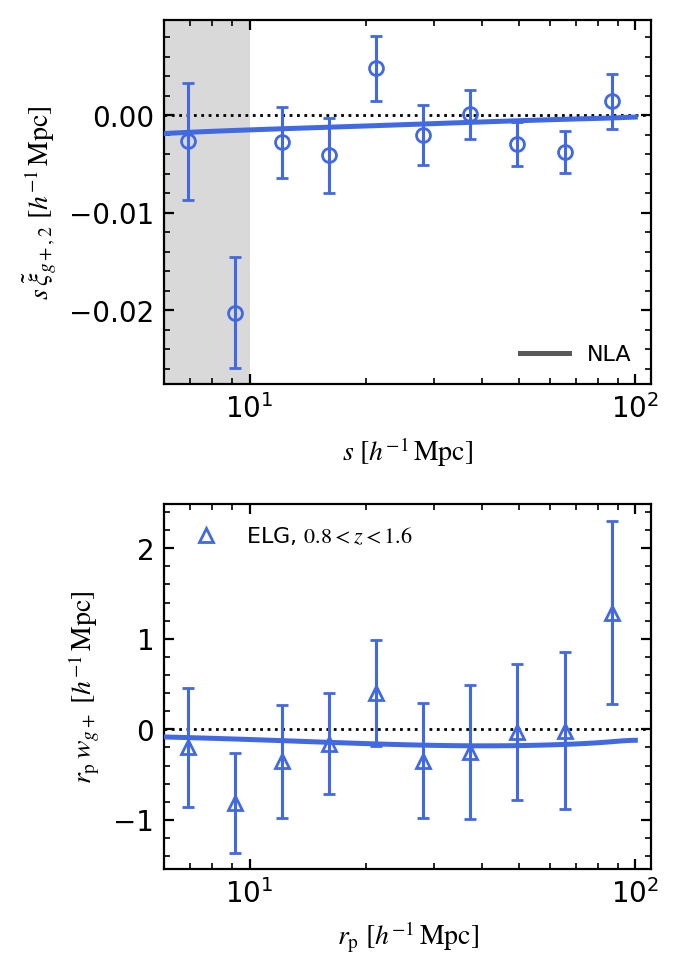}
   \caption{Same as Fig.~\ref{fig:corrfunc_blue} \redit{for the ELG sample, with the NLA model only.}}
   \label{fig:corrfunc_elg}
\end{figure}

\begin{table*}
\caption{Best-fit NLA model parameters for the \emph{red} galaxy samples (BGS red and LRG) from the multipole analysis ($\tilde\xi_{gg,0}$ and $\tilde\xi_{g+,2}$).}
\label{tab:fits_red}
\centering
\small
\setlength{\tabcolsep}{4pt}
\begin{tabular}{lcccccccc}
\hline\hline
Sample & $z$ range & $z_{\rm IA}$ & $\log M_\star$ & $N_{\rm shapes}$ & $\sigma_e$ & $\mathcal{R}$ & $b_1$ & $A_{\rm IA}^{\tilde\xi_{g+,2}}\ (\chi^2_\nu)$ \\
\hline
\multicolumn{9}{c}{\textit{BGS red}} \\
\hline
Global & $0.1$--$0.5$ & $0.25$ & $10.73 \pm 0.40$ & $583\,617$ & $0.236$ & $0.80$ & $1.390 \pm 0.040$ & $1.663 \pm 0.147\ (1.61)$ \\
$M_\star$ bin 1 & $0.1$--$0.5$ & $0.17$ & $10.03 \pm 0.13$ & $55\,434$ & $0.254$ & $0.82$ & -- & $2.211 \pm 0.424\ (1.87)$ \\
$M_\star$ bin 2 & $0.1$--$0.5$ & $0.20$ & $10.29 \pm 0.05$ & $55\,434$ & $0.258$ & $0.81$ & -- & $1.764 \pm 0.449\ (1.91)$ \\
$M_\star$ bin 3 & $0.1$--$0.5$ & $0.21$ & $10.44 \pm 0.04$ & $55\,433$ & $0.257$ & $0.80$ & -- & $1.597 \pm 0.492\ (1.39)$ \\
$M_\star$ bin 4 & $0.1$--$0.5$ & $0.23$ & $10.56 \pm 0.03$ & $55\,435$ & $0.253$ & $0.80$ & -- & $2.070 \pm 0.493\ (1.89)$ \\
$M_\star$ bin 5 & $0.1$--$0.5$ & $0.24$ & $10.67 \pm 0.03$ & $55\,434$ & $0.247$ & $0.80$ & -- & $1.113 \pm 0.482\ (1.17)$ \\
$M_\star$ bin 6 & $0.1$--$0.5$ & $0.25$ & $10.77 \pm 0.03$ & $55\,434$ & $0.240$ & $0.79$ & -- & $0.857 \pm 0.471\ (1.20)$ \\
$M_\star$ bin 7 & $0.1$--$0.5$ & $0.27$ & $10.86 \pm 0.03$ & $55\,433$ & $0.232$ & $0.79$ & -- & $1.474 \pm 0.448\ (2.13)$ \\
$M_\star$ bin 8 & $0.1$--$0.5$ & $0.28$ & $10.96 \pm 0.03$ & $55\,435$ & $0.222$ & $0.79$ & -- & $1.321 \pm 0.410\ (1.40)$ \\
$M_\star$ bin 9 & $0.1$--$0.5$ & $0.29$ & $11.07 \pm 0.03$ & $47\,565$ & $0.213$ & $0.79$ & -- & $1.076 \pm 0.451\ (1.64)$ \\
$M_\star$ bin 10 & $0.1$--$0.5$ & $0.30$ & $11.17 \pm 0.03$ & $37\,374$ & $0.204$ & $0.79$ & -- & $0.219 \pm 0.492\ (1.52)$ \\
$M_\star$ bin 11 & $0.1$--$0.5$ & $0.32$ & $11.29 \pm 0.04$ & $28\,783$ & $0.192$ & $0.79$ & -- & $2.872 \pm 0.476\ (1.26)$ \\
$M_\star$ bin 12 & $0.1$--$0.5$ & $0.33$ & $11.44 \pm 0.08$ & $20\,001$ & $0.170$ & $0.80$ & -- & $5.724 \pm 0.548\ (1.64)$ \\
\hline
\multicolumn{9}{c}{\textit{LRG}} \\
\hline
Global (low-$z$) & $0.4$--$0.75$ & $0.59$ & $11.15 \pm 0.16$ & $201\,654$ & $0.217$ & $0.81$ & $1.952 \pm 0.054$ & $1.168 \pm 0.353\ (1.45)$ \\
Global (high-$z$) & $0.75$--$1.1$ & $0.86$ & $11.13 \pm 0.19$ & $208\,369$ & $0.213$ & $0.74$ & $2.114 \pm 0.064$ & $2.945 \pm 0.382\ (0.64)$ \\
$M_\star$ bin 1 & $0.4$--$0.75$ & $0.60$ & $11.03 \pm 0.05$ & $55\,044$ & $0.227$ & $0.81$ & -- & $1.006 \pm 0.674\ (1.48)$ \\
$M_\star$ bin 2 & $0.4$--$0.75$ & $0.60$ & $11.12 \pm 0.02$ & $44\,841$ & $0.223$ & $0.81$ & -- & $-0.082 \pm 0.579\ (1.51)$ \\
$M_\star$ bin 3 & $0.4$--$0.75$ & $0.59$ & $11.19 \pm 0.02$ & $38\,050$ & $0.219$ & $0.82$ & -- & $0.982 \pm 0.652\ (1.55)$ \\
$M_\star$ bin 4 & $0.4$--$0.75$ & $0.59$ & $11.27 \pm 0.03$ & $31\,500$ & $0.210$ & $0.81$ & -- & $1.304 \pm 0.795\ (1.90)$ \\
$M_\star$ bin 5 & $0.4$--$0.75$ & $0.59$ & $11.41 \pm 0.10$ & $30\,000$ & $0.191$ & $0.80$ & -- & $2.921 \pm 0.722\ (1.65)$ \\
$M_\star$ bin 1 & $0.75$--$1.1$ & $0.86$ & $10.95 \pm 0.07$ & $58\,721$ & $0.245$ & $0.69$ & -- & $2.560 \pm 0.731\ (0.56)$ \\
$M_\star$ bin 2 & $0.75$--$1.1$ & $0.85$ & $11.09 \pm 0.03$ & $47\,709$ & $0.213$ & $0.75$ & -- & $1.941 \pm 0.688\ (0.74)$ \\
$M_\star$ bin 3 & $0.75$--$1.1$ & $0.86$ & $11.18 \pm 0.03$ & $38\,548$ & $0.204$ & $0.76$ & -- & $2.604 \pm 0.737\ (0.67)$ \\
$M_\star$ bin 4 & $0.75$--$1.1$ & $0.87$ & $11.28 \pm 0.03$ & $31\,098$ & $0.197$ & $0.77$ & -- & $3.915 \pm 0.938\ (0.96)$ \\
$M_\star$ bin 5 & $0.75$--$1.1$ & $0.88$ & $11.42 \pm 0.11$ & $30\,000$ & $0.195$ & $0.77$ & -- & $3.740 \pm 0.808\ (0.83)$ \\
\hline
\end{tabular}
\tablefoot{$\log M_\star$ is the median stellar mass in units of $\log(M/h^{-2}M_\odot)$, with the standard deviation of the distribution within each bin. $N_{\rm shapes}$ is the number of galaxies in the shape sample. $\sigma_e$ is the per-component calibrated shape noise, computed following the definition of \citet{Heymans2012} as in \citet{HervasPeters2024}. $\mathcal{R}$ is the effective shear response, the weighted average of the diagonal elements of the per-galaxy \texttt{Metacalibration} response matrix, following \citet{Siegel2025}. $z_{\rm IA}$ is the pair-weighted effective redshift of the density--shape cross-correlation, at which the galaxy--shape multipole is modelled; the clustering monopole is modelled at the effective redshift of the density sample, $0.35$ (BGS red), $0.60$ (LRG, low-$z$), $0.91$ (LRG, high-$z$). The density sample of each tracer and redshift bin is its full clustering catalogue (for BGS, all BGS galaxies, shared by the red and blue samples), with which the global shape sample and its stellar-mass sub-samples are cross-correlated. Its linear bias $b_1$ is constrained by the clustering monopole, common to all sub-samples, and is therefore quoted only for the global samples. $A_{\rm IA}^{\tilde\xi_{g+,2}}$ is the IA amplitude from the multipole fit, with the reduced $\chi^2$ of the joint $(\tilde\xi_{gg,0},\tilde\xi_{g+,2})$ fit in parentheses.}
\end{table*}

\begin{figure*}
   \centering
   \includegraphics[width=0.97\textwidth]{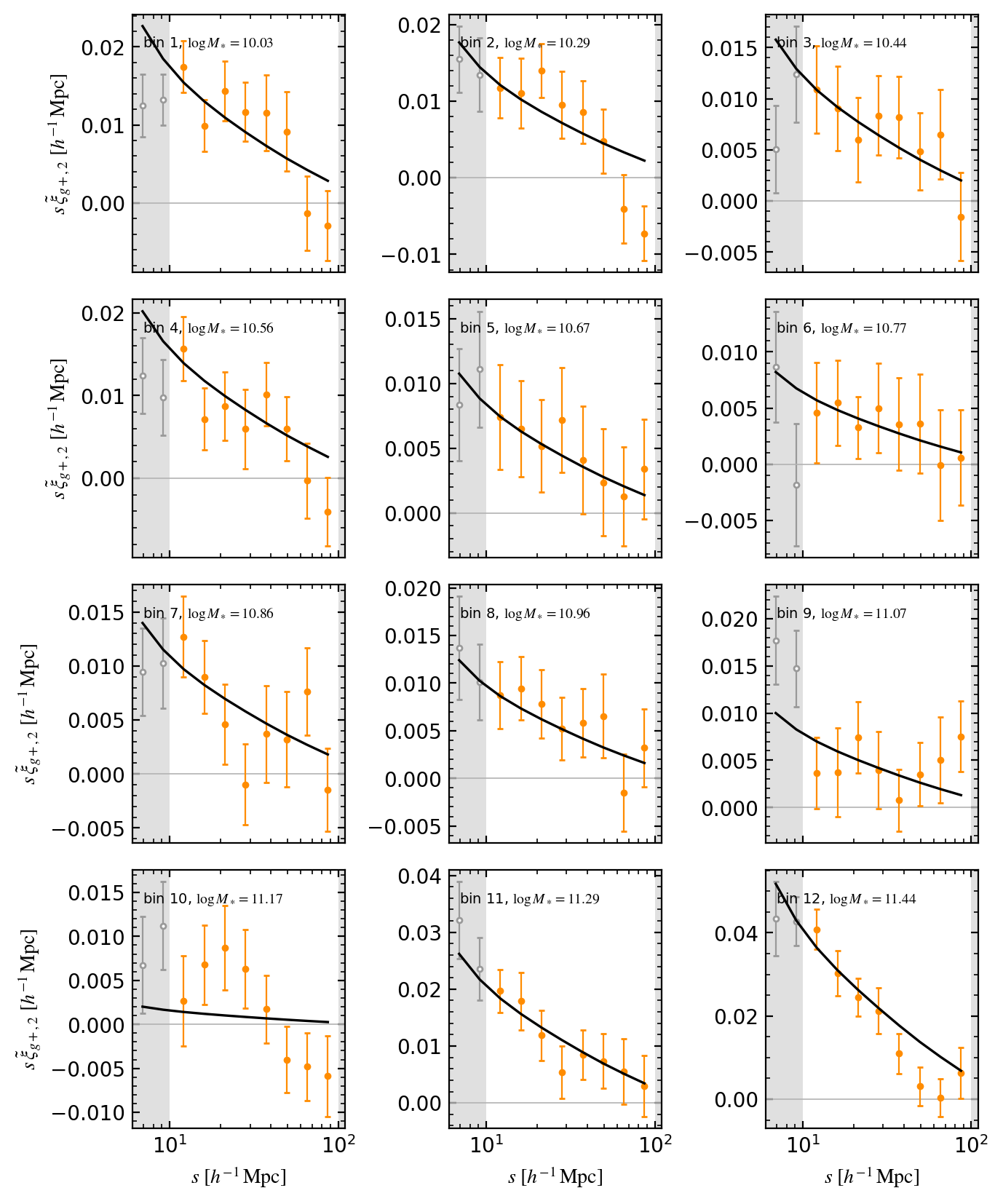}
   \caption{\cedit{Spin-2 quadrupole $s\,\tilde\xi_{g+,2}(s)$ for the twelve stellar-mass bins of the BGS red sample\redit{, numbered as in Table~\ref{tab:fits_red},} together with the best-fit NLA model of each bin (black line). Filled orange points are fitted ($10 < s < 100\,h^{-1}\,{\rm Mpc}$); \redit{the grey shading marks the scales excluded by the cut, where the} open grey points fall.}}
   \label{fig:bgs_snr_xi2}
\end{figure*}

\clearpage 
\section{Stellar-to-halo mass relation}
\label{app:shmr}

To interpret the mass dependence of the IA amplitude, we estimate the
stellar-to-halo mass relation (SHMR) of the BGS red sample from the
galaxy--galaxy lensing signal, using the conditional stellar mass function
(CSMF) halo-model formalism \citep{yang.etal08, vandenbosch.etal13,
cacciato.etal13}; our modelling pipeline is similar to that of
\citet{Dvornik2023}\cedit{, and is implemented in the public \textsc{NRVpy}
halo-model code\footnote{\url{https://github.com/rpaviot/NRV_HOD}, tag
\texttt{unions-desi-ia-paper}.}}. \cedit{In particular, we use the halo mass function and
halo bias of \citet{Tinker2010}, NFW density profiles \citep{NFW1997} with
the concentration--mass relation of \citet{Duffy2008}, and halo masses
defined within the radius enclosing 200 times the mean matter density
($M_{200{\rm m}}$), which is also the mass definition of the
\textsc{Dark Emulator} \citep{nishimichi.etal19} used for the non-linear
halo bias correction described below. The model is evaluated at the effective redshift of each
lens bin, in the cosmology of Sect.~\ref{sec:likelihood}, with halo masses
in $h^{-1}M_\odot$ and stellar masses in $h^{-2}M_\odot$.}

\begin{figure}
   \centering
   \includegraphics[width=\columnwidth]{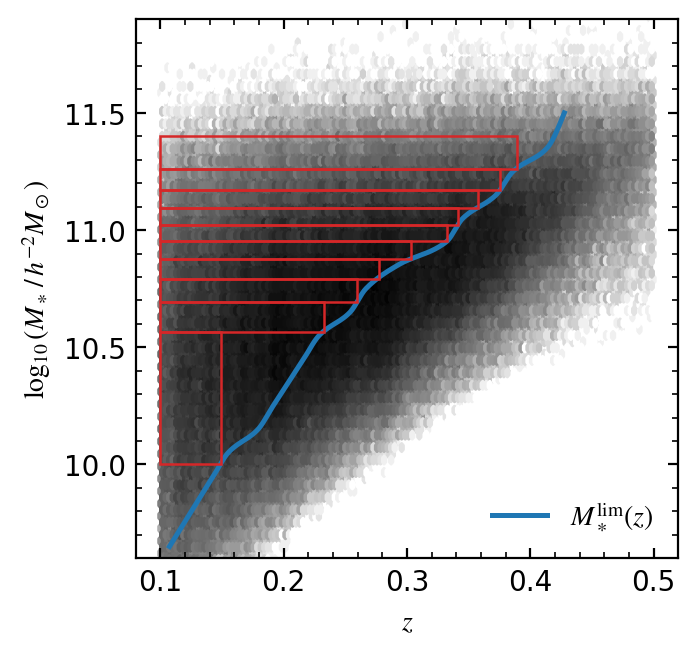}
   \caption{Definition of the volume-limited lens samples in the
   stellar-mass--redshift plane of the BGS red shape sample. The blue curve
   shows the stellar-mass completeness limit $M_\ast^{\rm lim}(z)$ derived
   with the turn-over method of \citet{wright.etal17}, and the red boxes the
   ten lens bins used in the fit, each truncated at the redshift where its
   lower mass edge crosses the completeness limit.}
   \label{fig:vlim_definition}
\end{figure}

\begin{figure}
   \centering
   \includegraphics[width=\columnwidth]{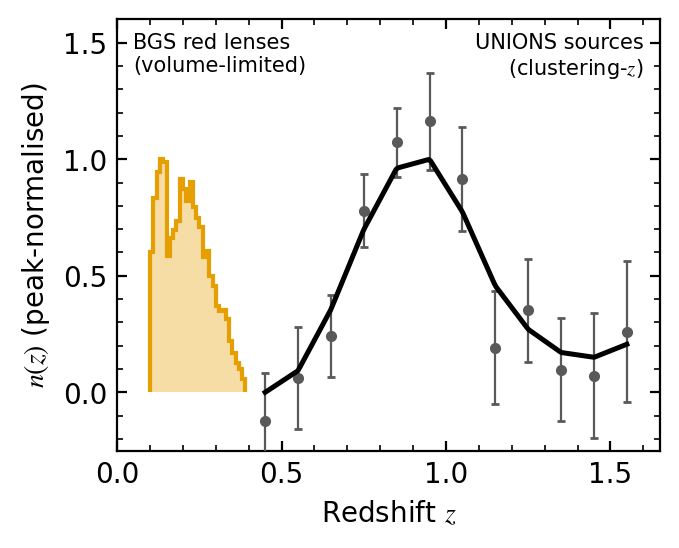}
   \caption{\redit{Redshift distributions of the galaxy--galaxy lensing samples, normalised to their peak: the volume-limited BGS red lens sample (all ten stellar-mass bins together) and the UNIONS sources ($0.70<z_{\rm B}<1.25$), whose distribution is measured with clustering redshifts (points with jackknife errors; the line is the smoothed, non-negative version used in Eq.~\eqref{eq:sigcrit_eff}).}}
   \label{fig:nz}
\end{figure}

Since the CSMF ties the galaxy abundance to halo mass, the lens samples must
be complete in stellar mass. We therefore construct volume-limited samples:
the limiting stellar mass of the red sample is estimated
as a function of redshift with the automated turn-over method of
\citet{wright.etal17} \citep[see also][]{Dvornik2023}\cedit{, applied here
along the line of sight: in thin stellar-mass slices, the completeness limit
is the comoving distance at which the number of galaxies per unit distance
turns over}, and the sample is
split into ten stellar-mass bins, each restricted to the redshift range
over which it is complete (Fig.~\ref{fig:vlim_definition}). The bin edges are
chosen to yield comparable forecast $\Delta\Sigma$ signal-to-noise per bin,
which narrows the bins at the high-mass end where the SHMR constraint
resides.

\cedit{We measure the excess surface mass density $\Delta\Sigma(r_{\rm p})$ around these ten lens bins (Sect.~\ref{sec:ggl}). The measurement uses the UNIONS shape catalogue \redit{\citep{HervasPeters2026}} as the source sample. \cedit{Unlike the IA shape sample, the sources are subject to the \redit{\textsc{ShapePipe}} quality selection, $0.5 < T/T_{\rm PSF} < 3$ (ratio of the galaxy to the PSF size\redit{; this lower bound is less conservative than the $0.707$ adopted for cosmic shear, \citealt{HervasPeters2026}}), a SExtractor windowed $r$-band magnitude ${\rm MAG\_WIN} > 20$, and $10 < {\rm S/N} < 200$\redit{, which mainly removes stars and spurious detections. We also require the sources to lie outside the bright-star and stellar-halo masks, a manual mask around large nearby galaxies (Messier and NGC objects), and the \textsc{MaxiMask} mask of image defects and cosmic rays \citep{Paillassa2020}, and correct them for PSF leakage \citep{Guerrini2025,HervasPeters2026}}. They are further selected in photometric redshift, $0.70 < z_{\rm B} < 1.25$, using the BPZ \redit{\citep{Benitez2000}} point estimate $z_{\rm B}$ from the UNIONS DR6 $ugriz$ photometry, so that the sources lie well behind the $z<0.5$ lenses. \redit{After masking, the source sample covers $4500\,{\rm deg}^2$ with an effective density of $2.7\,{\rm arcmin}^{-2}$.} \redit{These photometric redshifts are used here for the first time, and only to select the sources: their individual values never enter the measurement, whose source redshift distribution is calibrated with clustering redshifts, as described below.}} The excess surface mass density is estimated as a weighted sum over lens--source pairs,
\begin{equation}
   \Delta\Sigma(r_{\rm p}) = \frac{1}{\mathcal{R}}\left[
   \frac{\sum_{\rm ls} w_{\rm ls}\,\Sigma_{\rm crit,eff}(z_{\rm l})\,e_{\rm t,ls}}
        {\sum_{\rm ls} w_{\rm ls}}
   -
   \frac{\sum_{\rm rs} w_{\rm rs}\,\Sigma_{\rm crit,eff}(z_{\rm r})\,e_{\rm t,rs}}
        {\sum_{\rm rs} w_{\rm rs}}\right],
   \label{eq:dsigma_est}
\end{equation}
where the pair weight is $w_{\rm ls} = w_{\rm l}\,w_{\rm s}\,\Sigma_{\rm crit,eff}^{-2}(z_{\rm l})$, $e_{\rm t,ls}$ is the tangential ellipticity of the source relative to the lens\cedit{\footnote{The two estimators follow opposite sign conventions: $e_+$ is positive when the shape is aligned along the direction to the density tracer (radial alignment), whereas the lensing convention takes $e_{\rm t}$ positive for a tangential alignment around the lens.}}, $\mathcal{R}$ is the mean \texttt{Metacalibration} shear response, $w_{\rm l}$ is the lens weight \redit{($w_{\rm DESI}$, i.e.\ Eq.~\eqref{eq:desi_weight} without $w_{\rm FKP}$)}, and $w_{\rm s}=w_{\rm SP}$ is the source shape weight (Sect.~\ref{sec:unions}). The second term is the same weighted sum evaluated around the random-lens catalogue (index ${\rm rs}$, with $w_{\rm rs}$ and $z_{\rm r}$ defined analogously). The subtraction of the random signal decreases the variance at large scales \citep{Singh2017}. Rather than relying on individual source photometric redshifts, we compute an effective critical surface density for each lens redshift $z_{\rm l}$ as a weighted integral over the source redshift distribution, following \citet{Fortuna2025}:
\begin{equation}
   \Sigma_{\rm crit,eff}^{-1}(z_{\rm l}) = \int_{z_{\rm l}}^{\infty}\mathrm{d}z_{\rm s}\; n(z_{\rm s})\,\Sigma_{\rm crit}^{-1}(z_{\rm l},z_{\rm s})\,,
   \label{eq:sigcrit_eff}
\end{equation}
where $n(z_{\rm s})$ is the weighted redshift distribution of the sources. The latter is calibrated with clustering redshifts, obtained by cross-correlating the UNIONS sources with the DESI spectroscopic samples following the methodology of \citet{dAssignies2025}\cedit{, using the DESI DR1 LRG and ELG LSS catalogues as reference samples\redit{: the cross-correlations are measured at $1.5<r_{\rm p}<5\,h^{-1}\,{\rm Mpc}$ in slices of $\Delta z=0.1$ against the LRGs ($0.4<z<0.8$), the combined LRG and ELG sample ($0.8<z<1.1$) and the ELGs ($1.1<z<1.6$), and divided by the bias of each reference slice, obtained from its auto-correlation. The resulting distribution (Fig.~\ref{fig:nz}) has a mean redshift of $0.97$, with a jackknife uncertainty of $0.04$, and only $2\%$ of its weight lies below $z=0.6$}. The covariance of $\Delta\Sigma$ is estimated by jackknife over $100$ $K$-means regions of the lens footprint. We verified on the brightest lens bins that the boost factor is consistent with unity, as expected from the redshift separation between lenses and sources, and apply no boost or intrinsic-alignment correction; the lens magnification bias\redit{, estimated with the finite-difference method of \citet{Wenzl2024}, which we adapt to the DESI target selections by perturbing the total and fibre fluxes and the stellar masses of the lenses and re-applying the full sample selection,} is negligible for these lenses and is not subtracted. Distances are computed in the same Planck cosmology as the IA fits (Sect.~\ref{sec:likelihood}), and $\Delta\Sigma$ is expressed in $h\,M_\odot\,{\rm pc}^{-2}$ as a function of $r_{\rm p}$ in $h^{-1}\,{\rm Mpc}$.}}

In the model, the CSMF specifies the occupation of central and satellite
galaxies as a function of halo mass, and the halo-model prediction for
$\Delta\Sigma$ includes the baryonic contribution of the lens galaxies
themselves, the non-linear halo bias correction $\beta^{\rm NL}$
\citep{mead.verde21} computed with \textsc{Dark Emulator}, and baryonic
feedback through a rescaling of the
halo density-profile concentration \citep[Eq.~15 of][]{Dvornik2023}. The free
parameters are the four parameters of the central SHMR, the scatter of the
central relation, the slope and the two normalisation parameters of the
satellite CSMF, and the two profile rescaling amplitudes for matter and
satellites. \cedit{All parameters have flat priors, except the low-mass slope
of the SHMR, $\gamma_1$, which is not constrained by our data (the lens bins
lie above the knee of the relation) and for which we adopt a Gaussian prior
$\gamma_1\sim\mathcal{N}(7.10,\,2.0)$ centred on the value of
\citet{Dvornik2023}; the $\gamma_1$ of Table~\ref{tab:csmf_bestfit} is
therefore prior-dominated. The two profile rescaling amplitudes are bounded
above by unity, i.e.\ the matter and satellite profiles cannot be more
concentrated than the dark matter.} In addition to the ten lensing profiles,
the likelihood includes the measured comoving number density of each bin,
which anchors the abundance side of the relation. \cedit{The number density is
measured on the lens sample itself, \redit{with the weighting that reproduces the DR1 $n(z)$ tables released with} \citet{DESI2024II} and \citet{Ross2025}: the galaxies are
counted with their completeness, redshift-failure and imaging-systematic
weights (without the FKP weight), and the count is divided by the comoving
volume of the bin over the effective footprint area, i.e.\ the area traced
by the random catalogues weighted by the fraction of tiles observed at each
position. We verified that this procedure applied to the DESI LRG sample
reproduces the published DESI DR1 LRG $n(z)$. The uncertainty on the number
density is estimated by jackknife over the same $100$ regions as
$\Delta\Sigma$.}

\cedit{The quantity minimised is
\begin{align}
   \chi^2 \propto \sum_{i=1}^{10}\Big[
   &\left(\boldsymbol{\Delta\Sigma}_i - \boldsymbol{m}_i\right)^{\rm T}
   \mathbf{C}_i^{-1}
   \left(\boldsymbol{\Delta\Sigma}_i - \boldsymbol{m}_i\right) \nonumber\\
   &+ \frac{\left(n_{{\rm gal},i} - \hat n_{{\rm gal},i}\right)^2}{\sigma^2_{n_{{\rm gal},i}}}
   \Big]\,,
   \label{eq:chi2_csmf}
\end{align}
where, for each lens bin $i$, $\boldsymbol{\Delta\Sigma}_i$ and
$\boldsymbol{m}_i$ are the measured and model lensing profiles over the fitted
scales, $\mathbf{C}_i$ the corresponding jackknife covariance,
$n_{{\rm gal},i}$ and $\hat n_{{\rm gal},i}$ the measured and model number
densities. The covariance between different lens bins is neglected.} The fit is performed on
scales $0.5 < r_\mathrm{p} < 60\,h^{-1}\,\mathrm{Mpc}$\cedit{, i.e.\ ten
$\Delta\Sigma$ points and one number density per bin, $110$ data points for
$10$ free parameters. We sample the posterior with \textsc{nautilus}, with $2000$ live points and until the effective sample size
reaches $10\,000$, and quote in Table~\ref{tab:csmf_bestfit} the marginal
posterior medians with their $16$--$84\%$ intervals, together with the maximum
a posteriori (MAP) sample of the chain, which we use as the best-fit model in
what follows. The MAP model yields $\chi^2 = 142.2$ for $100$ degrees of
freedom: $116.4$ from the lensing profiles ($100$ points, i.e.\
$\chi^2_{\Delta\Sigma} = 116.4$ for $90$ degrees of freedom when the profiles
are considered alone) and $25.8$ from the number densities ($10$ points). A
direct minimisation with \textsc{Minuit} \citep{James1975} reaches \redit{nearly the same
minimum ($\chi^2 = 143.2$}, every parameter within $0.5\sigma$ of the MAP
sample) but pushes both profile rescaling amplitudes \redit{to or close to} their upper
bound and returns parabolic errors that misrepresent the posterior, which is
skewed along the $M_0$--$\gamma_2$ and $f_{\rm h}$--$f_{\rm s}$
degeneracies: the marginal medians sit about $1\sigma$ away from the mode,
and $f_{\rm h} = 0.83^{+0.12}_{-0.17}$ and $f_{\rm s} = 0.87^{+0.10}_{-0.17}$
are constrained below unity.}

\redit{In the MAP model, the satellite fraction of the lens bins decreases monotonically with stellar mass, from $26\%$ in the lowest-mass bin to $2\%$ in the most massive one. Evaluated over the stellar-mass ranges of the twelve BGS red alignment bins, it falls from $31\%$ to below $1\%$, the last bin being an extrapolation beyond the most massive lens bin. Since $M_{\rm eff}$ is defined from the central occupation alone (Eq.~\ref{eq:meff}), it describes the haloes of the central galaxies; the satellites, which reside in more massive host haloes, make up a quarter to a third of the lowest-mass alignment bins and less than $10\%$ above $\log(M_\star/h^{-2}M_\odot)\simeq11$.}

\cedit{\redit{We have also tested the fibre-assignment correction. Instead of the standard DESI completeness weights, we used individual-inverse-probability (IIP) weights, which up-weight each target that receives a fibre by the inverse of its fibre-assignment probability, which \citet{lange.etal24} showed to give an unbiased galaxy--galaxy lensing signal. We repeated the entire measurement and fit on the DESI DR1 LSS catalogues processed with these weights, keeping the lens binning, the source sample, the redshift distributions and the fitting configuration fixed, so that the completeness treatment is the only difference. The lensing profiles are unchanged within their uncertainties. The number densities, however, are lower, so the abundance anchor shifts the stellar-to-halo mass relation coherently towards higher halo masses. This moves the halo-mass scale of the $A_{\rm IA}$--$M_{\rm eff}$ relation of Sect.~\ref{sec:results}, but not its shape, and leaves our conclusions unchanged.}}

\begin{table}
\caption{CSMF parameters: posterior median with the $16$--$84\%$ interval, and the maximum a posteriori (MAP) sample of the \textsc{nautilus} chain.}
\label{tab:csmf_bestfit}
\centering
\setlength{\tabcolsep}{4pt}
\begin{tabular}{l l r r}
\hline\hline
Parameter & Description & Median & MAP \\
\hline
$\log_{10} M_0$ & SHMR normalisation & $10.72^{+0.07}_{-0.06}$ & $10.65$ \\
$\log_{10} M_1$ & Characteristic halo mass & $11.18^{+0.11}_{-0.09}$ & $11.11$ \\
$\gamma_1$ & SHMR low-mass slope & $9.1^{+1.4}_{-1.5}$ & $9.6$ \\
$\gamma_2$ & SHMR high-mass slope & $0.188^{+0.020}_{-0.025}$ & $0.208$ \\
$\sigma_{\rm c}$ & Central SHMR scatter & $0.248^{+0.007}_{-0.008}$ & $0.249$ \\
$\alpha_{\rm s}$ & Satellite CSMF slope & $-1.21^{+0.11}_{-0.09}$ & $-1.27$ \\
$b_0^{\rm s}$ & Satellite norm.\ intercept & $-0.41^{+0.17}_{-0.17}$ & $-0.55$ \\
$b_1^{\rm s}$ & Satellite norm.\ slope & $1.10^{+0.10}_{-0.11}$ & $1.18$ \\
$f_{\rm h}$ & Matter profile rescaling & $0.83^{+0.12}_{-0.17}$ & $0.92$ \\
$f_{\rm s}$ & Satellite profile rescaling & $0.87^{+0.10}_{-0.17}$ & $1.00$ \\
\hline
\end{tabular}
\tablefoot{$M_0$ is in $h^{-2}M_\odot$ and $M_1$ in $h^{-1}M_\odot$ ($M_{200{\rm m}}$). MAP: the maximum a posteriori sample of the chain, the best-fit model of Fig.~\ref{fig:deltasigma_bestfit}; $\chi^2 = 142.2$ there (Eq.~\eqref{eq:chi2_csmf}, 100 $\Delta\Sigma$ points and $10$ number densities for 10 free parameters).}
\end{table}

\begin{figure}
   \centering
   \includegraphics[width=\columnwidth]{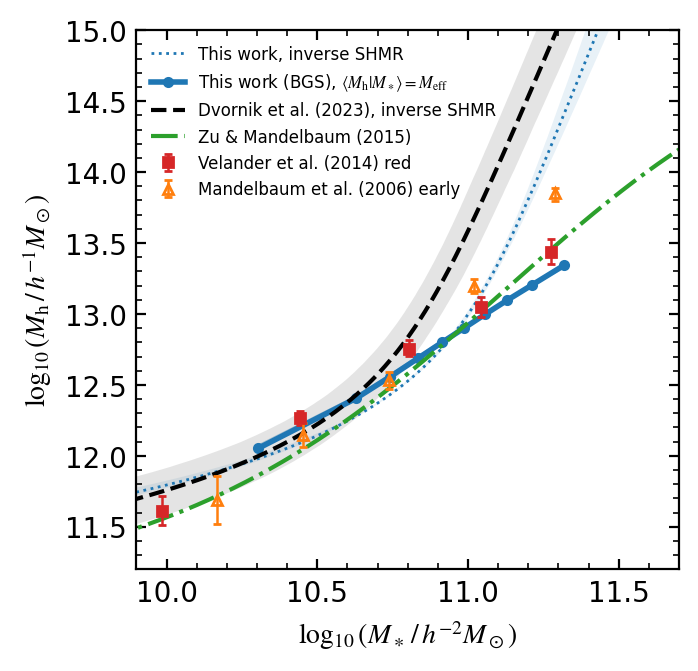}
   \caption{\redit{Mean halo mass of the central galaxies at fixed stellar mass for the BGS red lens bins (points and solid line, with the $16$--$84\%$ posterior band) and inverse of our central SHMR (dotted), compared with the inverse SHMR of \citet{Dvornik2023} (dashed, with its $16$--$84\%$ band) and the weak-lensing constraints compiled by \citet{zu.mandelbaum15}.}}
   \label{fig:shmr_literature}
\end{figure}

\begin{figure*}
   \centering
   \includegraphics[width=0.97\textwidth]{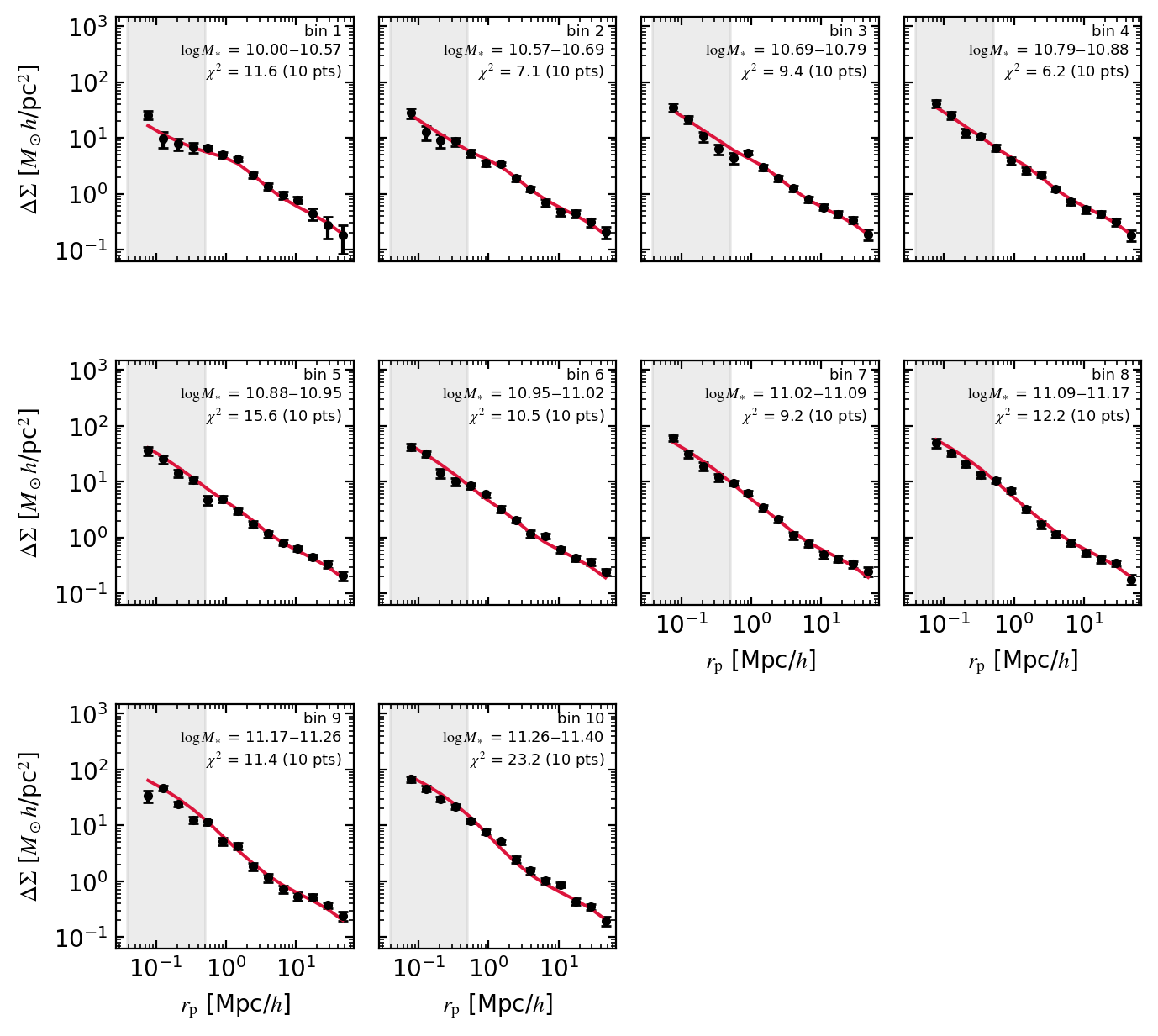}
   \caption{Excess surface density profiles of the ten volume-limited BGS red lens bins (points, jackknife errors) and the \cedit{MAP} CSMF halo model (red curves)\redit{; each panel gives the stellar-mass range of the bin and its contribution to the $\chi^2$ of Eq.~\eqref{eq:chi2_csmf}}. The shaded region marks the scales excluded from the fit.}
   \label{fig:deltasigma_bestfit}
\end{figure*}

The resulting central SHMR is shown in Fig.~\ref{fig:shmr_literature},
together with the relation of \citet{Dvornik2023} and the
weak-lensing constraints compiled by \citet{zu.mandelbaum15}. \redit{Over the stellar-mass range of our lens bins, $10.3\leq\log(M_\star/h^{-2}M_\odot)\leq11.3$, our $\langle M_{\rm h}|M_\star\rangle$ agrees with the relation of \citet{zu.mandelbaum15} to within $0.1$\,dex for $10.6\leq\log(M_\star/h^{-2}M_\odot)\leq11.15$, and lies up to $0.2$\,dex above it at the low-mass end and $0.2$\,dex below it at the high-mass end. It agrees with the points of \citet{velander.etal14} to within $0.1$\,dex up to $\log(M_\star/h^{-2}M_\odot)=11.0$ and lies $0.14$\,dex below their most massive point, while the early-type points of \citet{mandelbaum.etal06} lie $0.25$--$0.55$\,dex above our relation above $\log(M_\star/h^{-2}M_\odot)=11$.} The measured $\Delta\Sigma$ profiles
and the \cedit{MAP} model are shown in Fig.~\ref{fig:deltasigma_bestfit}.

\end{appendix}

\end{document}